\documentclass[%
superscriptaddress,
pra,
]{revtex4-2}

\usepackage{xcolor}
\usepackage{graphicx}
\usepackage{dcolumn}
\usepackage{bm}
\usepackage{braket}
\usepackage[normalem]{ulem}
\usepackage{amsmath}
\usepackage{siunitx}
\usepackage{upgreek}

\usepackage{multirow}

\begin{document}


\title{Coherence  of Microwave and Optical Qubit Levels in Neutral Thulium 
}

\author{D.\,Mishin}
\affiliation{P.N.\,Lebedev Physical Institute, Leninsky prospekt 53, 119991 Moscow, Russia}

\author{D.\,Tregubov}
\affiliation{P.N.\,Lebedev Physical Institute, Leninsky prospekt 53, 119991 Moscow, Russia}

\author{N.\,Kolachevsky}
\affiliation{P.N.\,Lebedev Physical Institute, Leninsky prospekt 53, 119991 Moscow, Russia}
\affiliation{Russian Quantum Center, Bolshoy Bulvar 30,\,bld.\,1, Skolkovo IC, 121205 Moscow, Russia}
\date{\today}

\author{A.\,Golovizin}       
\email{artem.golovizin@gmail.com}
\affiliation{P.N.\,Lebedev Physical Institute, Leninsky prospekt 53, 119991 Moscow, Russia}
\affiliation{Russian Quantum Center, Bolshoy Bulvar 30,\,bld.\,1, Skolkovo IC, 121205 Moscow, Russia}
\date{\today}

\begin{abstract}

Hyperfine-encoded qubits in alkali atoms have established themselves as robust platforms for quantum computing, while alkaline-earth-like elements expand the state manipulation toolbox through their rich spectrum of optical transitions and metastable states.
In this work, we demonstrate that thulium is a viable candidate for quantum computing, combining advantages of hyperfine qubit encoding with a rich energy-level structure of alkaline-earth-like atoms.
We describe protocols for the initial state preparation and state-selective readout, and show single-qubit operations on the microwave transition at $1 497$\,MHz.
We demonstrate ground state hyperfine qubit coherence times up to $T_2^* = 22^{+2}_{-2}$\,s and $T_2 = 55^{+59}_{-14}$\,s, representing record-scale performance for neutral-atom systems.
Furthermore, we show operations involving metastable optical states, including shelving for the state-selective readout as well as coherent population transfer of the ground state qubit with coherence time  primarily limited by the metastable level natural lifetime of $112$\,ms.
These results mark the first step toward using thulium for quantum computing applications and highlight its promising characteristics.

\end{abstract}

\maketitle

\section{Introduction}
\label{Section:Introduction}

Today there is no consensus yet on the leading platform for building a universal quantum computer: superconductors demonstrate potential for scalability \cite{kam2024characterization,edman2024hardware} and the shortest operation times \cite{anferov2024superconducting,howard2023implementing}, ions exhibit unmatched coherence times \cite{wang2021single} and high operation fidelity \cite{loschnauer2024scalable,srinivas2021high}, while neutral atoms offer thousand-scale atomic registers \cite{manetsch2024tweezer}, independent all-to-all connectivity achieved via atomic transport \cite{bluvstein2024logical,evered2023high}, along with high-fidelity single-qubit operations and two-qubit gate fidelity above 99.5\% \cite{ma2023high,evered2023high}.  
Alkali atoms, particularly Rb and Cs, with qubits encoded in the $m_F=0$ sublevels of the ground-state hyperfine doublet (exhibiting nearly infinite natural depolarization time $T_1$), remain the leading platform for neutral-atom quantum computing due to their simple energy-level structure, convenient transition wavelengths and large polarizability.
These atoms are also used in hybrid systems that utilize two or more different types of particles simultaneously \cite{sheng2022defect, singh2022dual,anand2024dual,fang2025interleaved} which significantly expands the toolkit for quantum simulations and error correction mechanisms.
In the meantime, alkaline-earth-like atoms (Yb and Sr) offer enhanced state manipulation capabilities through their two-valence-electron structure, providing: (i) kHz-linewidth optical transitions for deep laser cooling and selective readout \cite{norcia2023midcircuit,saskin2019narrow,urech2022narrow}; (ii) optical metastable states that can be used for temporal storage of ground-state qubit \cite{barnes2022assembly}, expansion of the Hilbert space and operations within the qudit framework \cite{jia2024architecture,huie2025three,nikolaeva2025scalable}, as well as single-photon Rydberg excitation for two-qubit gates \cite{jia2024architecture,ma2023high,shaw2025erasure}; (iii) Rydberg atoms confinement in optical tweezers \cite{wilson2022trapping}.
In these atoms, qubits are typically encoded in nuclear spin magnetic sublevels of the ground state, keeping them insensitive to most external perturbations.
However, this encoding requires application of a substantial bias magnetic field (tens to hundreds gauss) to create sufficient energy splitting for single- and two-qubit gates, which requires a high-precision control over the bias magnetic field noise.

\begin{figure*}[h]
\center{
\resizebox{0.5\textwidth}{!}{
\includegraphics{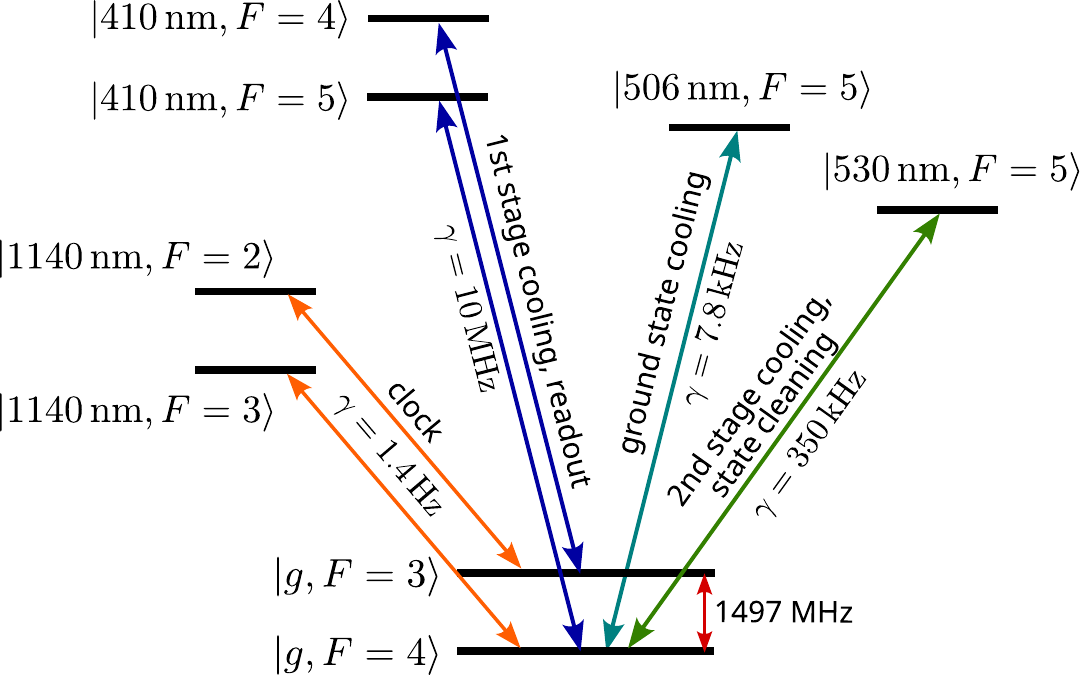}
}\\
\caption{Simplified thulium level structure.
The following transitions are of interest for experiments discussed in this work: 410\,nm transition is used for the first-stage laser cooling and population detection, 530\,nm transition serves for the second-stage laser cooling and polarization purification, and 506\,nm transition enables ground-state cooling in the optical lattice.
This work is focused on investigating the properties of the 1497\,MHz hyperfine qubit transition and 1140\,nm clock transitions to optical metastable states.
}
\label{fig:level structure}}
\end{figure*}

Here, we demonstrate that thulium atoms combine advantageous features of both aforementioned platforms: the rich electronic structure (see Fig.\,\ref{fig:level structure}) of alkaline-earth-like atoms and robust hyperfine-state (1\,497\,MHz splitting \cite{mishin2024combined}) qubit encoding analogous to Rb and Cs.
Thulium possesses a single stable isotope $^{169}$Tm featuring a nuclear spin $I=1/2$.
Its large ground-state magnetic moment of $4\,\mu_B$ ($\mu_B$ is the Bohr magneton), arising from a single vacancy in $4f$ electronic shell ($L=3$), makes thulium attractive for quantum simulations \cite{khlebnikov2019random, khlebnikov2021characterizing,davletov2020machine}. 
Simultaneously, the 1140\,nm transition between the fine-structure sublevels ($S=1/2$) of the ground state in thulium atoms  is used in optical clocks, benefiting from its exceptionally low sensitivity to black-body radiation \cite{golovizin2019inner, golovizin2021simultaneous,golovizin2021compact}.
The upper metastable level of this clock transition has a lifetime of 112\,ms, hence can be used for optical-metastable-ground (\textit{omg}) state architecture or qudit encoding schemes, as discussed above.

Previous works have established essential techniques for thulium atom manipulation, including multi-stage laser cooling \cite{sukachev2010sub,vishnyakova2014two,provorchenko2023deep}, sideband cooling to the ground vibrational state in optical lattice \cite{provorchenko2024laser}, identification of magic wavelengths (813\,nm and 1063.5\,nm) for clock transitions \cite{golovizin2019inner,mishin2022effect}, precise determination of both the ground-state hyperfine splitting value and the absolute clock transition frequency, with thorough characterization of their sensitivity to dominant external perturbations \cite{sukachev2016inner,golovizin2019inner,golovizin2021simultaneous,mishin2024combined}.

In this study, we investigate the properties of the qubit encoded in $m_F=0$ magnetic sublevels of the thulium ground state hyperfine structure.
Sections \ref{Section:Initial state preparation} and \ref{Section:Readout technique} provide a detailed description of the initial state preparation and readout protocols in the experiment.
Section \ref{Section: COHERENCE PROPERTIES OF THE GROUND STATE HYPERFINE TRANSITION} focuses on analyzing the coherence time of the hyperfine qubit using Ramsey spectroscopy and dynamical decoupling technique, while in Section \ref{Coherent operations with metastable states} we demonstrate coherent qubit transfer to $m_F=0$ sublevels of the optical metastable states.
Finally, in Section \ref{Section:conclusion}, we summarize the results, discuss possible improvements to the experimental setup, and outline the potential and future steps for using neutral thulium atoms in quantum information science.

\section{Preparation of polarization-pure initial state}
\label{Section:Initial state preparation}

Preparation of the initial state is a key stage of the experiment, as it directly affects both the fidelity of operations in quantum computing (part of state preparation and measurement ``SPAM'' errors \cite{ryan2021realization, jayakumar2024universal}) and the stability of measurements in quantum sensors and optical clocks (improves the signal-to-noise ratio).

In this work, we prepare atoms in one of the $\ket{g,F=4,m_F=0}$ or $\ket{g,F=3,m_F=0}$ states.
The same levels serve as initial states for synthetic frequency spectroscopy in thulium optical clock \cite{golovizin2021simultaneous} and previously, several methods based on the optical pumping technique were implemented to prepare atoms in these states \cite{fedorova2019optical,fedorova2020simultaneous}, but all resulted in significant atomic heating. 

\begin{figure*}[ht!]
\center{
\resizebox{\textwidth}{!}{
\includegraphics{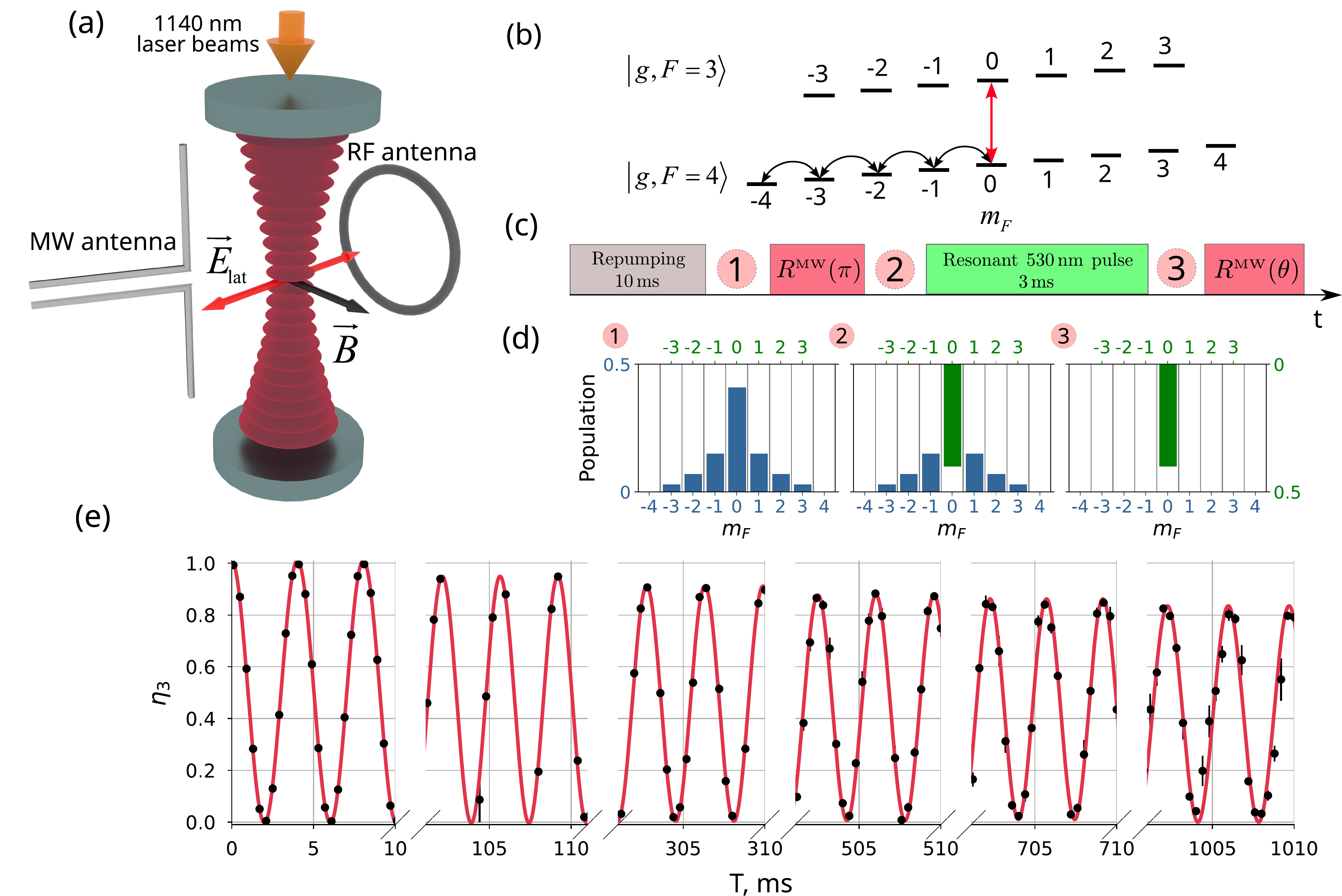}
}\\
\caption{Initial state preparation.
(a) Simplified scheme of the experimental setup.
The orientations of the $\sim1.5$\,GHz MW and $\sim 800$\,kHz RF antennas, 1063.5\,nm optical lattice polarization, and bias magnetic field are presented.
The 1140\,nm clock laser beam is aligned with the optical lattice.
(b) Diagram of hyperfine and magnetic sublevels of the ground state.
Transitions depicted with black arrows are used for RF repumping of atoms to the central magnetic sublevel.
The MW transition being studied in this paper is indicated by the red arrow.
(c) State preparation pulse sequence.
$R^{MW}$ indicates application of the MW pulse of the corresponding length ($\pi$-pulse with $\sim2$\,ms length, or a pulse of any necessary duration corresponding to $\theta$-angle rotation).
(d) Schematic histograms depicting atomic population distribution across magnetic sublevels of the $\ket{g,F=4}$ (blue) and  $\ket{g,F=3}$ (green) states at the times shown in the figure (c).
(e) Rabi oscillations on the MW transition are shown with the excitation pulse duration on the horizontal axis and the relative population of the $\ket{g,F=3}$ level $\eta_3$ on the vertical axis.
Error bars correspond to 1\,s.d. statistical uncertainty and for most experimental points are below the marker size.
The observed damping of the oscillations is fully described by atomic depolarization; for more details, see the Methods section.}
\label{fig:Repumping}}
\end{figure*}

Here, we use RF transitions to redistribute atoms between magnetic sublevels and hyperfine MW transition for selective $\ket{m_F=0}$ state shelving during polarization purification. 
Here and hereafter the experimental setup is similar to the one used in \cite{mishin2024combined,golovizin2021compact}, and its simplified scheme is shown in Fig.~\ref{fig:Repumping}(a) while RF and MW transitions used for the state preparation are shown in Fig.~\ref{fig:Repumping}(b).
Laser cooling of thulium is described in detail elsewhere \cite{provorchenko2024laser,vishnyakova2014two}.
After the final cooling stage, the atoms remain trapped in an optical lattice formed at a near-magic wavelength of $1063.5$\,nm \cite{golovizin2019inner, golovizin2021simultaneous} for the 1140\,nm clock transition.
In all experiments, the polarization of the lattice is perpendicular to the applied bias magnetic field and the lattice depth is set to $U_L=k_B\times5\,\mu\textrm{K}=100\,E_r$, where $E_r\approx h\times1$\,kHz is the recoil energy of the lattice photon.
The current coil (RF antenna) is driven by a signal from a direct digital generator at a frequency close to the splitting of the magnetic sublevels determined by the magnetic field applied in the experiment.
Microwave transition between the hyperfine levels is driven with a half-wave dipole antenna (MW antenna) adjusted to the ground state hyperfine splitting of 1497\,MHz \cite{mishin2024combined}, which is placed in close proximity to the CF100 window of the vacuum chamber \cite{golovizin2021compact}.
A maser-stabilised SRS RF Signal Generator SG382 is used as the signal source.
The maximum (and operational) microwave power on the antenna is 10\,dBm.
The built-in pulse modulation function combined with the use of an additional RF switch ZASWA-2-50DR+ reduces the signal level to less than $-120$\,dBm in ``off'' state, which is necessary for implementing Ramsey spectroscopy with a free evolution time on the order of tens of seconds.

The key idea of the state preparation scheme is similar to the one used in \cite{sinuco2019microwave, wynands2005atomic} with some modifications specific to thulium atoms and the problem at hand.
The pulse scheme of the experiment and corresponding population distribution among $m_F$ states are shown in Fig.\,\ref{fig:Repumping} (c) and (d) with the main steps described below:
\begin{enumerate}
  \item After the final stage of deep laser cooling at a wavelength of $506$\,nm, more than $99\%$ of atoms are in the ground vibrational state in the optical lattice and occupy the $\ket{g, F=4, m_F=-4}$ state \cite{provorchenko2024laser}.
  
  \item In the current experimental configuration, pumping of atoms to the central magnetic sublevel is performed using $5\,\text{ms}$  radio-frequency pulse.
  The RF frequency is swiped from $800\,\text{kHz}$ to $785\,\text{kHz}$ (for an applied magnetic field $B = 0.6\,\text{G}$) to sequentially transfer atoms between adjacent $m_F$ sublevels from $m_F=-4$ to $m_F=0$ (see  Fig.~\ref{fig:Repumping}(b)).  
  After this, about $40\,\%$ atoms occupy the central magnetic sublevel, while the remaining atoms are distributed across the other sublevels, as schematically shown in Fig.\,\ref{fig:Repumping}(d,1).
  \item Using a 2-ms microwave $\pi$-pulse, we transfer atoms from  $\ket{g,F=4,m_F=0}$ to $\ket{g,F=3, m_F=0}$.
  This transition is isolated from other $\ket{g,F=4,m_F}$ to $\ket{g,F=3, m'_F}$ transitions (the closest one is shifted by at least  60\,kHz) due to the bias magnetic field of $B_0=0.6$\,G applied in the experiment, Fig.\,\ref{fig:Repumping}(d,2).
  \item We remove atoms remaining in the ground $\ket{g, F=4}$ state using a $3\,\text{ms}$ pulse of $530\,\text{nm}$ radiation resonant with the second-stage laser cooling transition with a natural linewidth of $350\,\text{kHz}$.  
  Since the difference in the hyperfine splittings of the upper and ground levels of this transition is $614\,\text{MHz}$, atoms in the $\ket{g, F=3, m_F=0}$ state are not excited, and thus there is no depolarization or losses from this state.  
  After this pulse, only the $\ket{g, F=3, m_F=0}$ sublevel remains populated as depicted in Fig.\,\ref{fig:Repumping}(d,3) (see polarization purity characterization in Methods).
  \item Now, we can prepare any desired population distribution between the $\ket{g, F=4, m_F=0}$ and $\ket{g, F=3, m_F=0}$ sublevels by selecting the appropriate length of the second microwave pulse. 
  The Fig.\,\ref{fig:Repumping}(e) illustrates the probability of detecting atoms in the $\ket{g, F=3, m_F=0}$ state as a function of microwave pulse duration.
  Over 250 periods of coherent oscillations are observed, with their amplitude decaying due to depolarization of atoms in one of the states (see Methods section for details).
  The initial few oscillation periods show the possibility to prepare any target state with a purity greater than $99(1)\%$, with uncertainty currently constrained by the precision of the readout calibration procedure.

\end{enumerate}

In the current experimental configuration, more than half of the atoms are lost during the state cleaning (step 3) after the non-ideal pumping process from the outermost to the central magnetic sublevel.
The demonstrated state preparation efficiency of approximately $40\%$ of atoms, all occupying the ground vibrational state in the optical lattice, is sufficient for the current research and optical clock performance.
Further improvements could be achieved through coherent population transfer using microwave pulse sequences, see Methods for details.

\section{Single-cycle state-selective readout}
\label{Section:Readout technique}

For a careful analysis of the characteristics of transitions between central magnetic sublevels, we need to independently detect the populations of $\ket{g,F=4,m_F=0}$ and $\ket{g,F=3,m_F=0}$ states.
This can be achieved through minor modifications to the method used in thulium optical lattice clock experiments \cite{golovizin2021simultaneous}.

\begin{figure*}[h]
\center{
\resizebox{0.5\textwidth}{!}{
\includegraphics{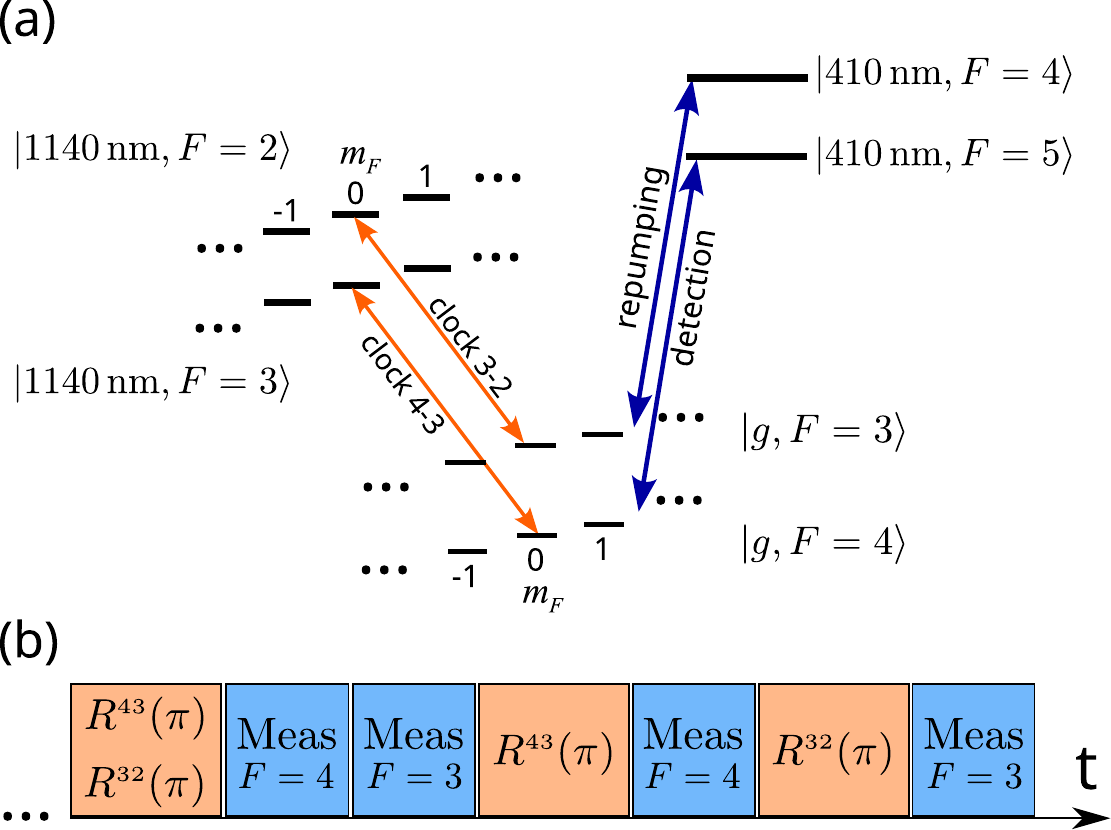}
}\\
\caption{Readout scheme.
(a) Thulium level structure with marked transitions used in the readout procedure.
(b) The readout pulse sequence. 
$R^{FF'}(\pi)$ indicates application of $\pi$-pulse on $\ket{g, F,m_F=0} \rightarrow\ket{1140\,\textrm{nm}, F',m_F=0}$ transition with a length of $1$\,ms.
``Meas'' corresponds to the detection of atom number in the specified $F$ state, see text for details.
Each measurement includes $400$\,\textmu s $410$\,nm probe pulse radiation and $4$\,ms dead time to let the detected atoms leave the spectroscopy region.
}
\label{fig:readout technique}}
\end{figure*}

To detect atoms in $\ket{g, F=4}$ state, we use a pulse of probe radiation resonant with the $\ket{g,F=4}\rightarrow\ket{410\,\textrm{nm},F=5}$ transition. Here $\ket{410\,\textrm{nm}}$ denotes the level used for the first stage laser cooling at a wavelength of $\lambda_{b}=410$\,nm with $\Gamma_b = 2\pi\times10$\,MHz, see Fig.\,\ref{fig:readout technique}(a).
This allows us to get the highest scattering photon rate among any other transitions used in the experiment. 
This readout is destructive — atoms that scatter the probe radiation heat up and exit the spectroscopy region within $\sim0.5$\,ms; in the current experiment, a waiting time of $4$\,ms is set between consecutive measurements.
As a result, after each measurement, the population of the $\ket{g,F=4}$ state is depleted to zero.

After measuring the $\ket{g, F=4}$ population, we can detect atoms in $\ket{g, F=3}$ state.
To do that we repump these atoms to the $\ket{g, F=4}$ state via the $\ket{g, F=3} \rightarrow \ket{410\,\textrm{nm}, F=4}$ transition, which takes approximately $50$\,\textmu s, and repeat the readout procedure described above.  

To determine the central magnetic sublevel populations, we use a shelving technique involving metastable states, see Fig.\,\ref{fig:readout technique}(b).
Immediately before the readout procedure, we transfer atoms from both $\ket{g,F=4,m_F=0}$ and $\ket{g,F=3,m_F=0}$ central magnetic sublevels into metastable states using $\pi$-pulses of optical radiation at a wavelength of 1140\,nm.
Since the 410\,nm radiation does not interact with atoms in these states, during the readout we sequentially obtain the total population of the nonzero magnetic sublevels ${N}_{4,m_F\ne0}$ and ${N}_{3,m_F\ne0}$.
We next return the atoms from the metastable levels using corresponding $\pi$-pulses of optical radiation and measure their number using the method described above, thereby obtaining the populations of the central magnetic sublevels ${N}_{4,m_F=0}$ and ${N}_{3,m_F=0}$.

This technique can be used to detect specific erasure errors \cite{wu2022erasure,ma2023high}, particularly those arising from spontaneous decay outside the qubit subspace.
In general, it additionally enables selective readout of populations in other magnetic sublevels, which is especially useful for systems with extended internal level structure, such as qudit-based architectures.

It is worth noting that the probe radiation for the $\ket{g, F=4} \rightarrow \ket{410\,\textrm{nm}, F=5}$ transition, while being $\sim360$\,MHz detuned from the $\ket{g, F=3} \rightarrow \ket{410\,\textrm{nm}, F=4}$ transition, induces off-resonant scattering.
At the same time, population transfer using 1140\,nm radiation in the current setup is not perfect and metastable states have a finite lifetime of $112$\,ms \cite{golovizin2019inner} which should be accounted for.
All these effects depend on the experimental configuration but remain constant over time, and can therefore be calibrated, as detailed in Methods.
We also note that these effects can be significantly suppressed by switching to narrowband readout and minimizing back reflections of the 1140\,nm radiation, which is also discussed in detail in the Methods.

\section{COHERENCE PROPERTIES OF THE GROUND STATE HYPERFINE TRANSITION}
\label{Section: COHERENCE PROPERTIES OF THE GROUND STATE HYPERFINE TRANSITION}

Having described the pure initial state preparation protocol, microwave transition driving technique, and state-selective readout, we now investigate the coherence properties of the $m_F=0$ sublevels of the ground state hyperfine structure using Ramsey and dynamical decoupling sequences.

\subsection{Atom lifetime and depolarization}
\label{Subsection: Atom lifetime and depolarization}

The measurement time is ultimately limited by the lifetime of the atoms in the optical lattice.
In addition to collisions with the background gas, two- or many-particle collisions may also lead to losses.
The latter is strongly pronounced in magnetic atoms due to long-range dipole-dipole interaction (DDI): DDI causes spin flips, which result in depolarization and losses.
Thereby, the dynamics of the number of atoms is described by Eq.\,(\ref{eq:two_body_loss}) \cite{weiner1999experiments}
\begin{equation} \label{eq:two_body_loss}
   N(t) = \frac{N_0\cdot e^{-t/\tau}}{1+ ({\beta}/V)\tau\,N_0 (1-e^{-t/\tau})},  
\end{equation}
where $\tau$ is the exponential single-atom lifetime (due to collisions with the background gas), $N_0$ is the initial number of atoms, $V$ is the effective volume of the atomic cloud and ${\beta}$ is the two-body loss coefficient. 
The trapping volume is constant in our experiments and, assuming a Gaussian distribution of atoms along each axis, is equal to $V=(2\pi)^{3/2}\times0.16\,\textrm{mm}\times 0.16\,\textrm{mm}\times 0.4\,\textrm{mm} \approx 0.16\,\textrm{mm}^3$ with a total uncertainty of about $30\%$ coming from the camera calibration, possible atomic cloud expansion during the detection procedure and cloud size fluctuations.
Here the population dynamics of atoms in different states is studied to determine the maximum measurement time in the experiment.

\begin{figure*}[ht!]
\center{
\resizebox{0.9\textwidth}{!}{
\includegraphics{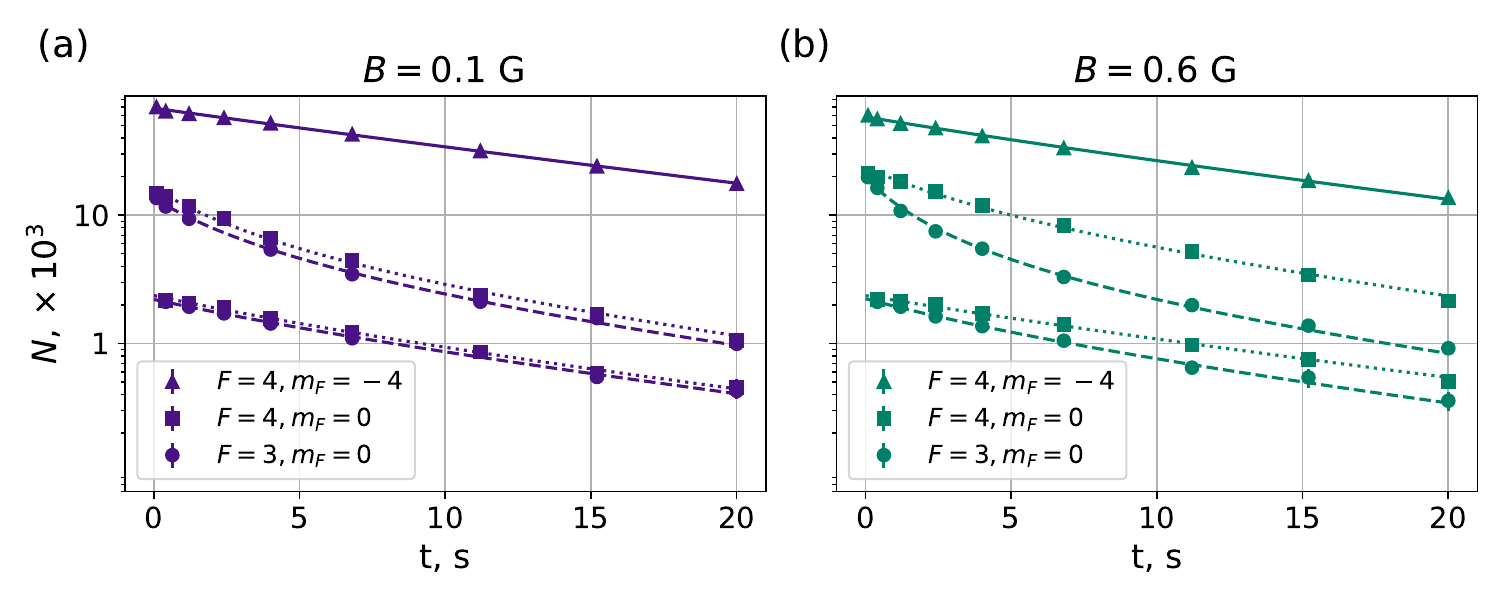}
}\\
\caption{Lifeteme measurements.
Time evolution of the number of atoms initially prepared in $\ket{F=4,m_F=-4}$ (triangle markers), $\ket{F=4,m_F=0}$ (square markers) and $\ket{F=3,m_F=0}$ (round markers) states is presented for magnetic fields of $B=0.1$\,G (a) and $B=0.6$\,G (b).
For $m_F=0$ states results are shown for different initial numbers of atoms.
Error bars correspond to 1\,s.d. statistical uncertainty and for most experimental points here are below the marker size.
The data is fitted with Eq.~(\ref{eq:two_body_loss}) (lines).}
\label{fig:lifetime}}
\end{figure*}

First, we investigate the lifetime of the atoms in the lowest-energy $\ket{g, F=4, m_F=-4}$ state, which is obtained directly after the deep laser cooling stage. 
Results for two different bias magnetic fields $B=0.1$\,G and $B=0.6$\,G are shown with triangle markers in Fig.\,\ref{fig:lifetime}(a) and (b), respectively.
Solid lines are the fits to the data with Eq.~(\ref{eq:two_body_loss}).
In both cases, the dependence is close to exponential, with a characteristic single-particle lifetime of $\tau=16.4(1.7)$\,s and two-body loss coefficients of $\beta(m_F=-4,B=0.1\,\textrm{G})=2.5(2.6)\times10^{-11}\,\textrm{cm}^3/\textrm{s}$ and $\beta(m_F=-4,B=0.1\,\textrm{G})=6.6(3.3)\times10^{-11}\,\textrm{cm}^3/\textrm{s}$.
The increase in the two-body loss coefficient at higher magnetic field may be associated with Feshbach resonances \cite{khlebnikov2019random}.
In subsequent measurements, we fix the obtained single-particle lifetime value $\tau=16.4$\,s.

Next, we repeat the experiment for atoms prepared in $\ket{g, F=4, m_F=0}$ and $\ket{g, F=3, m_F=0}$ states (square and circle markers in Fig.\,\ref{fig:lifetime}, respectively), with varying initial atom numbers differing by approximately one order of magnitude. 
A significant increase in the role of two-body interactions is observed: for these measurements, the two-body loss coefficients are substantially non-zero (see Table\,\ref{tab:lifetime}).
We note here that for the $F=4$ level, we observe atoms redistributing among non-zero magnetic sublevels (depolarization) without leaving the trapping region.
However, the developed readout method allows us to distinguish between loss and depolarization effects.
The temporal evolution of populations of $m_F=0$ and $m_F\ne0$ magnetic sublevels is examined in more detail in the Methods section.

\begin{table*}[ht!]
\caption{\label{tab:lifetime}   
Two-body loss rate for atoms prepared in different initial states.
}
\renewcommand{\arraystretch}{1.5}
\begin{ruledtabular}
\resizebox{1\linewidth}{\height}{
\begin{tabular}{lcccc}
                    &        &$\ket{F=4,m_F=-4}$    &$\ket{F=4,m_F=0}$   & $\ket{F=3,m_F=0}$ 
\\ \hline
\multirow{2}{*}{$\beta,\,\textrm{cm}^3/s$}
    &$B=0.1$\,G & $2.5(2.6)\times10^{-11}$&$2.8(0.8)\times10^{-9}$& $3.2(0.9)\times10^{-9}$       \\ \cline{2-5}
    &$B=0.6$\,G & $6.6(3.3)\times10^{-11}$& $1.1(0.3)\times10^{-9}$&$4.3(1.3)\times10^{-9}$\\ 
\end{tabular}
}
\end{ruledtabular}
\end{table*}

In our case, observed loss rates, when considered alongside the measurement threshold of $\sim20$ atoms (due to CMOS camera noise) lead to the measurement time limit of approximately $20$\,s.
Increasing the initial number of atoms enhances the role of two-particle interactions, which not only counteract the atom number increase by the end of the holding time but could also affect the coherence characteristics analyzed in subsequent sections.
Accordingly, following experiments are conducted in a regime where the initial atom number does not exceed $5\times10^3$, ensuring that two-body interactions remain sufficiently weak.
We note here that when working with individually trapped atoms in optical tweezers, both losses and depolarization caused by DDI will be strongly suppressed.

\subsection{Ramsey spectroscopy}
\label{Subsection: Ramsey spectroscopy}

\begin{figure*}[ht!]
\center{
\resizebox{0.7\textwidth}{!}{
\includegraphics{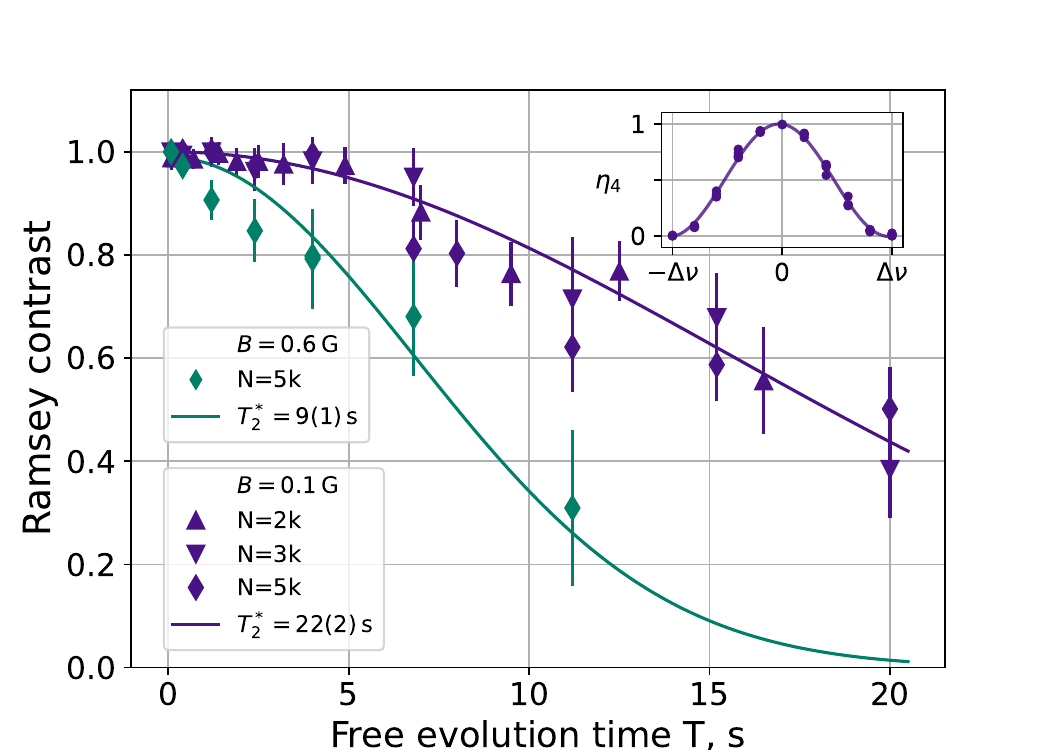}
}\\
\caption{
Ramsey contrast measurements.
Detected dependence of the Ramsey fringe contrast is presented for two magnetic field values $B=0.1$\,G (purple) and $B=0.6$\,G (green).
For $B=0.1$\,G experiment was conducted for different initial number of atoms in range 2000-5000 (different marker shapes).
Error bars represent 1\,s.d. uncertainty from the Ramsey fringes fit.
Solid lines show fit to the data  with Gaussian decay function Eq.\,(\ref{eq: gaussian decay}) with corresponding decay times $T_2^*$ of $22(2)$\,s and $9(1)$\,s for $B=0.1$\,G and $B=0.6$\,G respectively.
The inset shows an example of a typical Ramsey fringe pattern obtained in experiment, presenting the dependence of the relative population $\eta_4$ on the frequency detuning of the microwave radiation for a free evolution time of $T = 80$\,ms at $B=0.1$\,G.
}

\label{fig:Ramsey}}
\end{figure*}

We begin investigating the coherence properties of the $m_F=0$ sublevels of the ground-state hyperfine doublet with microwave Ramsey spectroscopy.
Here and henceforth, unless stated otherwise, we will be interested only in the number of atoms remaining in the central sublevels $N_{4,0}$ and $N_{3,0}$, where $N_{F,m_F}$ denotes the measured number of atoms in the sublevel $\ket{g,F,m_F}$.
The experimental sequence is the following: we prepare atoms in $\ket{F=3, m_F=0}$ state and apply two Ramsey MW $\pi/2$ pulses which are separated by variable free-evolution time $T$. 
For each $T$, we scan the frequency of the microwave radiation in the range of $\Delta\nu=1/T$  near the resonance in order to record Ramsey fringes.
Example can be seen in the inset in Fig.~\ref{fig:Ramsey}, where we depict the relative population 
\begin{equation}
    \eta_{4}=N_{4,0}/(N_{4,0} + N_{3,0})
\end{equation}
of $\ket{F=4,m_F=0}$ ground hyperfine level for $T=80$\,ms.
We fit this data with 
\begin{equation}
    \eta_4(\Delta\nu) = A + C/2 \times \cos(\pi T \Delta\nu + \phi_0)
    \label{eq:Ramsey fringes}
\end{equation}
with free parameters $A$ (offset), $C$ (contrast), and $\phi_0$ (initial phase of the Ramsey oscillations). 
Purple and green points in Fig.\,\ref{fig:Ramsey} show inferred contrast $C(T)$ for two bias magnetic fields $0.1$\,G and $0.6$\,G, respectively.

For $B=0.1$\,G the experiment was carried out with different initial atom numbers in the range from 2,000 to 5,000 (different point shapes in Fig.~\ref{fig:Ramsey}).
No significant changes in coherence time were observed within this range, and the solid purple curve illustrates the fit of all data with a Gaussian decay function 
\begin{equation}
    C(T) = C_0e^{-(T/T_2^*)^2}
    \label{eq: gaussian decay}
\end{equation}
with a $T_2^*(B=0.1\,\textrm{G}) = 22(2)$\,s.
Experimental data at $B=0.6$\,G is also fitted with Eq.\,(\ref{eq: gaussian decay}), giving $T_2^*(B=0.6\,\textrm{G}) = 9(1)$\,s, but the agreement is relatively poor. 
We attribute both the decrease of the $T_2^*$ and worse fitting at larger bias magnetic field to a larger influence of the magnetic field fluctuation.
The dependence of the hyperfine transition frequency between $m_F=0$ sublevel on the magnetic field is quadratic with coefficient $\gamma_\textrm{qz} = 852\,\textrm{Hz/G}^2$ \cite{mishin2024combined}, which corresponds to 8.5\,Hz and 306.7\,Hz shift at $B=0.1$\,G and $B=0.6$\,G, respectively.
This indicates much higher sensitivity to the temporal magnetic field fluctuations and inhomogeneities for higher $B$. 
This assumption is further supported by a more detailed analysis of Ramsey fringes, as described in the Methods section.

\subsection{Dynamical decoupling}
\label{Subsection: Dynamical decoupling}

A standard method for suppressing the influence of external fields on a quantum system is dynamical decoupling \cite{souza2012robust, biercuk2009optimized, khodjasteh2005fault}.
This technique uses intermediate $\pi$-pulses to mitigate decoherence effects arising from spatial inhomogeneities and temporal fluctuations of experimental conditions.
Increasing the number of these pulses broadens the frequency range of noise that can be effectively compensated through decoupling.

\begin{figure*}[h]
\center{
\resizebox{0.8\textwidth}{!}{
\includegraphics{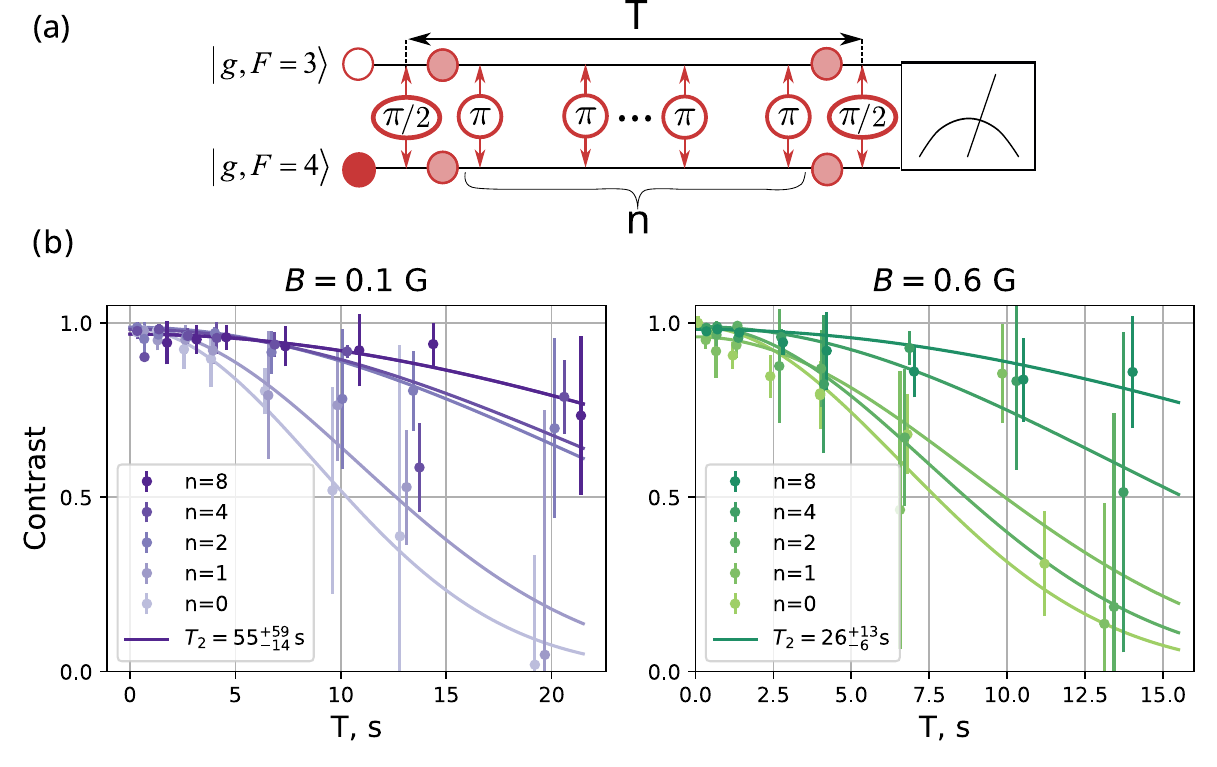}
}\\
\caption{The dynamical decoupling experiment.
(a) Measurement sequence: we start with all atoms prepared in $\ket{g,F=4,m_f=0}$ state; than we apply MW $\pi/2$-pulse followed by $n$ decoupling $\pi$-pulses; before the population measurement we apply second $\pi/2$ pulse to map state coherence into the population distribution. 
(b) Experimental results for magnetic fields $B=0.1$\,G and $B=0.6$\,G.
The data for different numbers $n$ of intermediate $\pi$-pulses are shown in various colors, where $n=0$ corresponds to the Ramsey  and $n=1$ to the Hahn echo measurement schemes.
Error bars represent 1\,s.d. statistical uncertainty.
Solid lines illustrate Gaussian decay fit results.
}
\label{fig:Decoupling}}
\end{figure*}

In order to estimate the coherence time $T_2$ which is less sensitive to external conditions, we perform a Carr-Purcell (CP) \cite{carr1954effects} dynamical decoupling sequence, see Fig.~\ref{fig:Decoupling}(a). 
Measurements are performed for $B=0.1$\,G and $B=0.6$\,G with varying numbers of intermediate pulses in the CP sequence, including $n=0$ (Ramsey scheme), $n=1$ (Hahn Echo scheme), as well as $n=2,4,8$.
The estimated contrast (see Methods for details) of the dynamical decoupling experiment is shown in Fig.\,\ref{fig:Decoupling}(b), with different colors corresponding to the different numbers of intermediate $\pi$-pulses. 
Colored solid lines show fit to the corresponding data with Gaussian decay function Eq.\,(\ref{eq: gaussian decay}).

For both magnetic fields, an increase in the number of intermediate $\pi$-pulses leads to a gradual increase in coherence time up to $T_2(B=0.1\,\textrm{G})=55^{+59}_{-14}$\,s and $T_2(B=0.6\,\textrm{G})=26^{+13}_{-6}$\,s, both for $n=8$.
We emphasize that these values give a lower bound on the achievable coherence time since (i) the measurement time is limited to $\sim20$\,s which is less than observed coherence times, (ii) the number of $\pi$-pulses can be increased further, and (iii) our measurement procedure provides a conservative estimate of contrast (see Methods).   
However, achieved estimates are already close to the record values demonstrated in neutral atoms \cite{young2020half, norcia2019seconds,tian2024extending, barnes2022assembly, manetsch2024tweezer}.
It is also worth noting that similar or even faster decay of the contrast for the Hahn echo ($n=1$) sequence compared to the Ramsey ($n=0$) scheme at $B=0.6$\,G supports the assumption of the leading role of the temporal fluctuations of the magnetic field rather than of some inhomogeneities over the atomic ensemble.

\section{ Coherence preservation during ground-metastable operations}
\label{Coherent operations with metastable states}

Finally, we demonstrate coherent manipulation of the atomic state between the ground and metastable levels. 
As we discussed in Introduction, this significantly broadens the toolkit to store, process, and measure quantum states.

\begin{figure*}[ht!]
\center{
\resizebox{0.9\textwidth}{!}{
\includegraphics{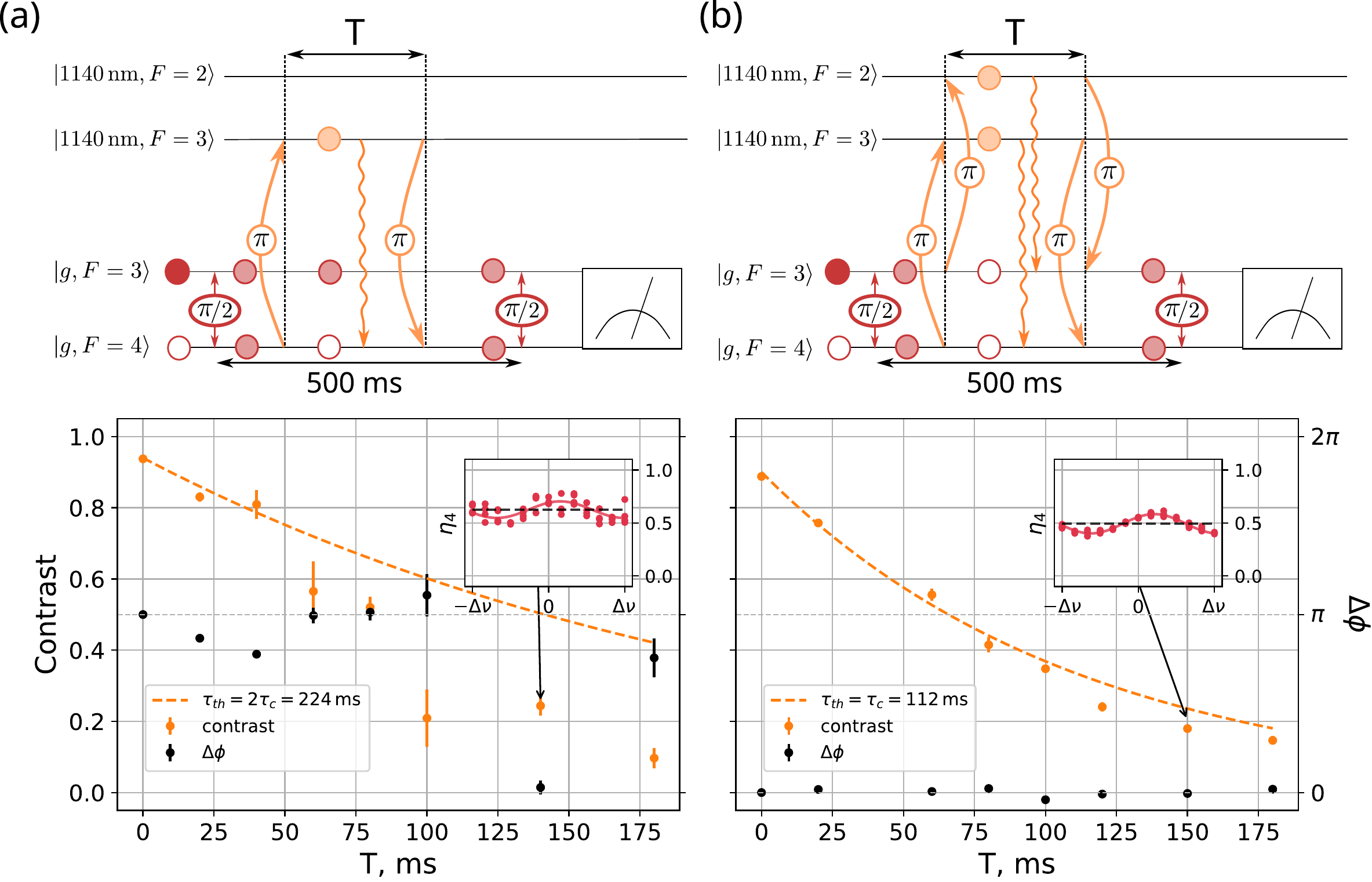}
}\\
\caption{
Measurements of the coherence time of the optical metastable states in (a) single- and (b) double-transition scheme.
The top row illustrates simplified measurement sequence.
We perform modified Ramsey measurement with $500$\,ms free evolution time: in the middle between two MW $\pi/2$-pulses we transfer the population of one (a) or both (b) of the qubit states to the optical metastable levels for a varying time $T$ and return them back to the ground state with corresponding $\pi$-pulses of $1140$\,nm radiation. 
Graphs in the bottom row show the experimental results.
Orange points illustrate contrast (left axis) inferred from Ramsey fringes analysis (see insets), and black points correspond to the phase shift $\Delta\phi$ (right axis) compared to the pulse sequence without intermediate optical pulses.
Error bars represent 1\,s.d.  uncertainty from Ramsey fringes fit.
Dashed lines illustrate the theoretical contrast decay rate limit, which equals $2\tau_c$ for the excitation of a single 1140\,nm transition and $\tau_c$ for bicolor interaction, where $\tau_c=112$\,ms is the optical metastable level lifetime.
}
\label{fig:Clock coherence}}
\end{figure*}

In order to do this, we insert optical manipulations between the microwave Ramsey spectroscopy pulses (see schematics in Fig.\,\ref{fig:Clock coherence}.)
We operate at $B=0.1$\,G  and set the time between two microwave $\pi/2$-pulses to be equal to 0.5\,s, which without intermediate optical pulses provides clean Ramsey fringes with the contrast above $99\%$ (see results in Sec.~\ref{Subsection: Ramsey spectroscopy}).
For each configuration of optical pulses we measure Ramsey fringes (see insets in Fig.\,\ref{fig:Clock coherence}) and infer their contrast and phase shift $\Delta\phi$ relative to the pure microwave sequence.
In these measurements we calculate contrast based on the total number of atoms detected in $\ket{g,F=4}$ and $\ket{g,F=3}$ states (before we considered only $m_F=0$ sublevels' populations) since spontaneous decay from the metastable states populates $m_F=0,\pm1$ ground state magnetic sublevels.

First, we investigate the performance of a ``single-transition'' scheme, when we optically excite atoms from only one of the ground levels, see Fig.\,\ref{fig:Clock coherence}(a). 
We apply two optical $\pi$-pulses on $\ket{g,F=4,m_F=0} \rightarrow \ket{1140\,\textrm{nm},F=3,m_F=0}$ transition, which are separated by time $T$ up to 180\,ms.
For $T=0$ we observe the expected phase accumulation of $\pi$ for the MW Ramsey fringes due to the applied optical $2\pi$-pulse, while the contrast is reduced to $0.94$ because of non-perfect efficiency of the 1140\,nm excitation (see Methods for details).
Dashed orange line shows expected exponential decrease of the contrast with $T_2^* = 2\tau_c$ ($\tau_c=112$\,ms is the optical metastable level lifetime).
One can clearly see that for $T\gtrsim50$\,ms deviation from this curve increases, which we attribute to limited coherence of the 1140\,nm laser.
This also leads to fluctuations of the imprinted phase shift reaching $\delta(\Delta\phi) \sim \pi$ at $T\gtrsim100$\,ms.

Next, we switch to ``double-transition'' configuration.
Here we utilize a simultaneous bicolor excitation scheme, which was originally developed for thulium optical clock \cite{golovizin2021simultaneous}.
Both $\ket{g,F=4,m_F=0} \rightarrow \ket{1140\,\textrm{nm},F=3,m_F=0}$ and $\ket{g,F=3,m_F=0} \rightarrow \ket{1140\,\textrm{nm},F=2,m_F=0}$ transitions are driven by radiation from a single laser source, with independent frequency control provided by two acousto-optic modulators, which eliminates the laser phase noise from the differential frequency of these two pulses. 
The measurement scheme, inferred MW Ramsey contrast and phase shift are shown in Fig.\,\ref{fig:Clock coherence}(b).
The major difference from a ``single-transition'' scheme is the stable imprinted phase shift (black points). 
It can be kept near zero by setting the correct frequency difference between two laser pulses matching the energy difference of the ground and metastable hyperfine states.
One can also see from the insets in Fig.\,\ref{fig:Clock coherence}, that MW Ramsey fringes are less noisy in this configuration compared to the ``single-transition'' scheme as the laser phase noise is eliminated.
The residual fluctuations of the accumulated phase we associate with the slow drift of the bias magnetic field, which we discussed above.
The MW Ramsey contrast, while starting from a lower value of $0.9$ at $T=0$ due to the doubled error of two 1140\,nm transitions excitation, is close to the exponential decay (dashed line) with $T_2=\tau_c$, which limits us in the bicolor interrogation scheme. 
We observe slightly lower than expected contrast at $T\gtrsim100$\,ms, which could be associated with the magnetic field fluctuations or relative phase deviations of two optical pulses due to fluctuations in non-common sections of the optical path.

\section{Discussion and Conclusion}
\label{Section:conclusion}
In this work, we have characterized the coherence properties of thulium atoms.
The 1497\,MHz transition was used to drive single-qubit rotations between $m_F=0$ sublevels of the ground state hyperfine structure.
We have demonstrated initial state preparation and selective readout protocols for this system.
Using Ramsey spectroscopy, we have measured the hyperfine qubit coherence time of $T_2^*=22_{-2}^{+2}$\,s, which extends to $T_2=55_{-14}^{+59}$\,s when implementing $n=8$ decoupling $\pi$-pulses.
These values are comparable to the record coherence times achieved in neutral atom systems \cite{young2020half, norcia2019seconds,tian2024extending, barnes2022assembly, manetsch2024tweezer}.
We have shown microwave qubit operations with a $\pi$-pulse fidelity exceeding $99(1)\%$ (with uncertainty limited by the current readout protocol) and visibility of more than 0.8 after 250 Rabi oscillations. 

We have also demonstrated the coherent population transfer from either or both of hyperfine qubit states simultaneously to the optical metastable levels via a 1140\,nm transition.
When working in a single-transition regime, the coherence time was constrained by the laser characteristics.
In contrast, the dual-transition scheme demonstrated effective laser noise cancellation, with coherence time limited by the optical metastable state natural lifetime of 112\,ms.
In this work we have already used state shelving in optical metastable states to enable state-selective readout, and this approach could be further adapted for ancilla-based operations, qudit encoding or ``\textit{omg}'' architecture protocols. 
Moreover, transfer of the ground state qubit to metastable state before the readout allows to detect population outside of qubit states, which can be used for error detection and correction \cite{wu2022erasure}.

Most of the limitations, which we faced during this work, are straightforwardly solvable. First, the readout procedure could be significantly improved with the use of a single-photon level camera (i.e. sCMOS, qCMOS or EMCCD).
This would enable single-atom resolution and fluorescence detection on a 530\,nm transition with a natural linewidth of 350\,kHz.
Such an approach could provide non-destructive readout, avoiding off-resonance excitation of the second qubit level.

Second, the influence of the magnetic field fluctuation, which currently limits achieved coherence times, can be significantly reduced by installing passive magnetic shields, deploying its active stabilization, or extending dynamical decoupling.
Reported results are obtained on an experimental setup designed for a compact thulium optical clock, and as its synthetic frequency is insensitive to Zeeman shift, it was not equipped with any dedicated systems.

Third, the preparation of atoms in the initial $m_F=0$ state could be done with close to unity efficiency by utilizing microwave transfer pulses. 
Here, the limiting factors are magnetic field noise (discussed above) and the power of the microwave radiation, which could be increased with a specially designed antenna and amplifier \cite{pershin2020microwave}.

Fourth, limited efficiency of population transfer to the optical metastable states is now determined by the 1140\,nm radiation reflection from the optical lattice enhancement cavity mirror and the necessity to align the clock beam with the optical lattice axis to work in the Lamb-Dicke regime.
The use of optical tweezers for individual addressing provides a natural solution to this problem.
Losses and depolarization due to DDI would also be suppressed when trapping single atoms into an optical tweezers array.  

It is worth noting that the optical tweezers could potentially reduce coherence times due to the significantly deeper trapping potential $U_T\sim k_B\times100\,\mu$K, comparing to $U_{L}\approx k_B\times5\,\mu$K of the optical lattice in the experiment.
However, as demonstrated in \cite{mishin2024combined}, the differential polarizability of thulium qubit levels at 1063.5\,nm trapping wavelength is on the order of $10^{-5}$ in atomic units with a ground-state scalar polarizability of 152\,a.u. \cite{mishin2022effect}.
Their ratio is three orders of magnitude lower than the typical value for cesium atoms (with demonstrated 12.6\,s coherence time) \cite{manetsch2024tweezer}, suggesting that the optical tweezer’s effect on thulium coherence properties should be substantially lower.
Moreover, this influence can be further decreased by deep laser cooling \cite{provorchenko2024laser}.
We also note that the currently employed magic wavelength of 1063.5\,nm falls within the operational range of commercially available high-power continuous-wave laser sources (such as fiber amplifiers exceeding 100\,W), enabling large-scale tweezer array formation.

In addition to solving these technical issues, realization of a universal quantum computer based on thulium requires  implementation of two-qubit operations, which can be achieved using the standard Rydberg blockade method.
Similar to alkali atoms, we plan to use two-photon excitation scheme from the ground state. 
Initial analysis showed that one can use $\ket{g} \rightarrow \ket{4f^{13}(^2F^\mathrm{o}_{7/2})6s6p(^1P^\mathrm{o}_{1})} \rightarrow \ket{4f^{13}(^2F^\mathrm{o}_{7/2})6s ns(nd)}$ transitions with the wavelengths of 409.5\,nm and $\sim 393$\,nm, which in counter propagating configuration allows almost full cancellation of the recoil effect.
We plan to experimentally investigate this in the near future, as well as the possibility of trapping a thulium atom in Rydberg state in the tweezer.

The demonstrated long coherence times and high-fidelity manipulations within the hyperfine structure of thulium’s ground and metastable optical states highlight its potential for quantum computing, simulation, and quantum memory applications.
Thulium uniquely combines the advantages of hyperfine qubit of alkali atoms with the versatile state control of alkaline-earth-like atoms. 
Furthermore, the spectral proximity of thulium and ytterbium cooling transitions simplifies dual-species operation, analogous to the demonstrated rubidium-cesium hybrid platform \cite{sheng2022defect, singh2022dual,anand2024dual,fang2025interleaved}, allowing to combine strengths of each qubit type.

\section{Methods}
\label{Section:Methods}

\subsection{General overview of the experimental setup}
\label{General overview of the experimental setup}

The current experimental setup is described in most detail in the works \cite{golovizin2021compact, golovizin2022control}.
Here, we  consider only a few of its aspects that are relevant to the experiments conducted in this study.

The characteristic magnetic field gradient for MOT is on the order of $10$\,G/cm and is generated using current coils in an anti-Helmholtz configuration.
Additional coils allow for the creation of a necessary field of up to $1$\,G along each axis (limited by the current sources), which is used to generate the bias field in the experiment and to compensate for laboratory and Earth's magnetic fields.
The characteristic field settling time is less than $100$\,ms for the values used in the experiment, limiting the minimum atom preparation cycle time to about $600$\,ms.
This includes Zeeman slowing, two stages of MOT, recapture of atoms into the optical lattice, and laser cooling down to the ground vibrational state.

In the current version of the system, instead of a classical Zeeman slower, we utilize magnetic field generated by MOT coils and an additional current coil, which creates an extra magnetic field gradient along the axis of propagation of the hot atoms from the atomic oven.
A counter-propagating laser beam resonant with the $\ket{g,F=4}\rightarrow\ket{410\,\textrm{nm},F=5}$ transition is focused on the output aperture of the atomic oven.
The low sensitivity of the clock transition frequency in thulium atoms to thermal radiation from the environment has allowed the atomic oven to be placed just $10$\,cm from the trapping and spectroscopy region.
This significantly reduces the system's size and is promising for the development of transportable optical clock \cite{golovizin2021compact}.
However, in some experiments, it may limit the lifetime of the trapped atoms due to degradation of the vacuum level in the system.

To mitigate this effect, a dual-chamber vacuum system with a narrow-line MOT loading procedure has been developed and demonstrated in the laboratory, allowing the separation of the hot atomic beam formation region from the spectroscopy region, and increasing the lifetime of atoms \cite{yaushev2025loading}.
In the future, this experimental configuration is planned to be used for further research of the group.

Atom detection is performed using an acA2040-55um Basler camera and an objective with a numerical aperture of approximately $0.2$.
At present, this configuration sets the minimum operational number of atoms at around 500; for signal levels lower than $\sim20$ atoms, camera noise significantly affects the measurement accuracy, which is crucial at low populations of one of the investigated states.

\subsection{Readout characterization}
\label{Subection:Readout characterization}

\subsubsection{Hyperfine states readout crosstalk}
\label{Subection:Hyperfine states readout crosstalk}
Beyond the constraints associated with the camera's sensitivity, the present readout scheme necessitates calibration to account for the effects of the readout pulses on the collected data.
As we discussed in the main text, we first detect the number of atoms on $\ket{g,F=4}$ level using resonant radiation on $\ket{g,F=4} \rightarrow \ket{410\,\textrm{nm}, F=5}$ transition, then we wait for 4\,ms in order for these atoms, which are heated by resonant radiation, to leave the detection volume.
To detect the number of atoms in $\ket{g,F=3}$, we repump them to $\ket{g,F=4}$ using $\ket{g,F=3} \rightarrow \ket{410\,\textrm{nm}, F=4}$ transition in less than 50\,\textmu s and repeat the detection procedure. 
Since the frequency difference between the transitions used for the detection $\ket{g,F=4} \rightarrow \ket{410\,\textrm{nm}, F=5}$ and repumping $\ket{g,F=3} \rightarrow \ket{410\,\textrm{nm}, F=4}$ is $360$\,MHz, or $36 \gamma_{410}$, the first detection pulse causes weak repumping from $\ket{g,F=3}$ level that leads to (i) an increase in the measured number of atoms in $\ket{g,F=4}$, and (ii) a decrease in the measured number of atoms in $\ket{g,F=3}$.

\begin{figure*}[ht!]
\center{
\resizebox{0.5\textwidth}{!}{
\includegraphics{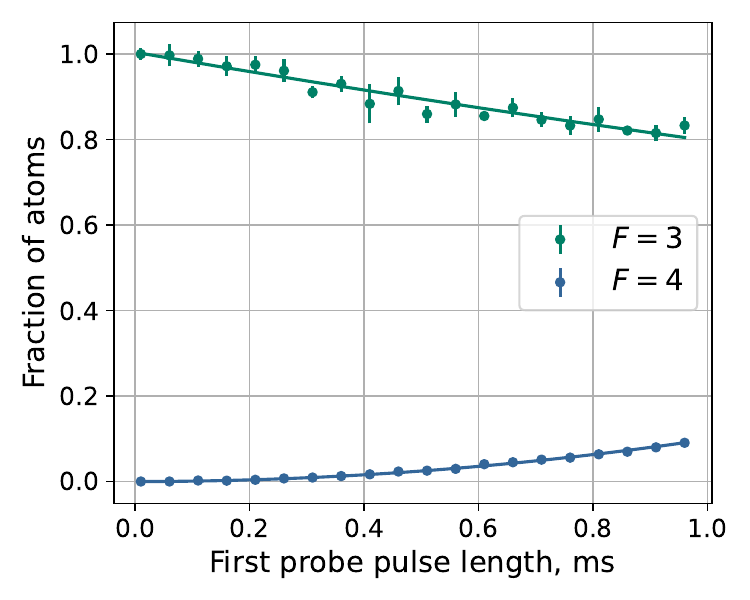}
}\\
\caption{Readout calibration.
Raw number of atoms detected during $\tilde{N}_{4}$ and $\tilde{N}_{3}$ readout steps with atoms prepared in $\ket{F=3}$ state.
Error bars correspond to 1\,s.d. statistical uncertainty.
Data is fitted with exponential decay and quadratic growth functions, see text.
}
\label{fig:Readout}}
\end{figure*}

For the characterization of this effect, we prepare atoms in the $\ket{g,F=3}$ state, as described in Sec.~\ref{Section:Initial state preparation}.
In this situation, the number of atoms at the $\ket{g,F=4}$ level can be considered zero, since the measurements are performed immediately after state cleaning, and in the absence of step (3) of state preparation, the signal is zero within the measurement accuracy.
Next, we measured the dependence of the number of atoms detected at the $\ket{g,F=4}$ and $\ket{g,F=3}$ level (using a second probe pulse of constant duration) on the duration of the first probe pulse.
The results are presented in Fig.~\ref{fig:Readout}.
Since losses occur due to heating of the atoms, the number of atoms in the $\ket{g,F=3}$ state was fitted using an exponential decay function with a characteristic time of $T = 4$\,ms (green solid line in Fig.~\ref{fig:Readout}).

To estimate the signal contributed by atoms from the $\ket{g,F=3}$ level when detecting atoms at the $\ket{g,F=4}$ level, we construct a simple model: the probability $P$ of an atom at the $\ket{g,F=3}$ level to interact with the probe radiation is assumed to be constant, and upon interaction, these atoms are "pumped" to the $\ket{g,F=4}$ level, where they start contributing to the signal throughout the entire duration of the probe illumination.
Thus, for small $P\tau$ parameter, the signal can be expressed as
\begin{equation}
    S(\tau)\propto\int_0^\tau n(t)\,dt=\int_0^\tau\left( 1-e^{-Pt}\right)\,dt=\tau-\frac{1-e^{-P\tau}}{P}\propto \tau^2,
\end{equation}
and we fit $\ket{g,F=4}$ data with a parabolic growth function (blue solid line in Fig.~\ref{fig:Readout}).

In the current setup, the readout pulse length is set to $0.4$\,ms as a compromise between the signal level and readout errors requiring corrections.
According to the obtained results, under experimental conditions, during the readout of atoms from the $\ket{g,F=4}$ level, approximately $1.5\%$ of the atoms residing at the $\ket{F=3}$ level contribute to the signal.
At the same time, due to the first readout pulse, the signal of $\ket{F=3}$ readout is underestimated by approximately $8.5\%$.
The intensity of the probe pulse was set to $I=2I_{sat}$ for all the experiments in this work, where $I_{sat}=180$\,\textmu$\textrm{W}/\textrm{mm}^2$ is the saturation intensity for the probe transition.
It is also worth noting that atoms in the metastable states are not subject to this effect.

It should also be noted that in order to ensure that the detected atoms leave the spectroscopy region and do not contribute to subsequent measurements, it is necessary to provide a flight time of at least $0.5$\,ms.
In the current configuration, this is achieved by introducing a dead time of approximately 4\,ms between consecutive readout cycles described in Sec.\,\ref{Section:Readout technique}.
This ensures suppression of the signal from non-dispersed atoms to a level well below $10^{-3}$ of the detected value, which is significantly smaller than the single-measurement error.

\subsubsection{1140\,nm transition calibration}
\label{Subection:Clock transition Rabi oscillations}

To calibrate the readout scheme with 1140\,nm pulses, it is necessary to account for the non-ideality of the transition excitation, as well as the lifetime of the metastable levels.
The latter was done using modeling via QuTiP and allows calculating the fractions of atoms that decayed from each of the metastable states into each of the ground states by the time of each detection event shown in Fig.~\ref{fig:readout technique}(b). 

\begin{figure*}[ht!]
\center{
\resizebox{0.5\textwidth}{!}{
\includegraphics{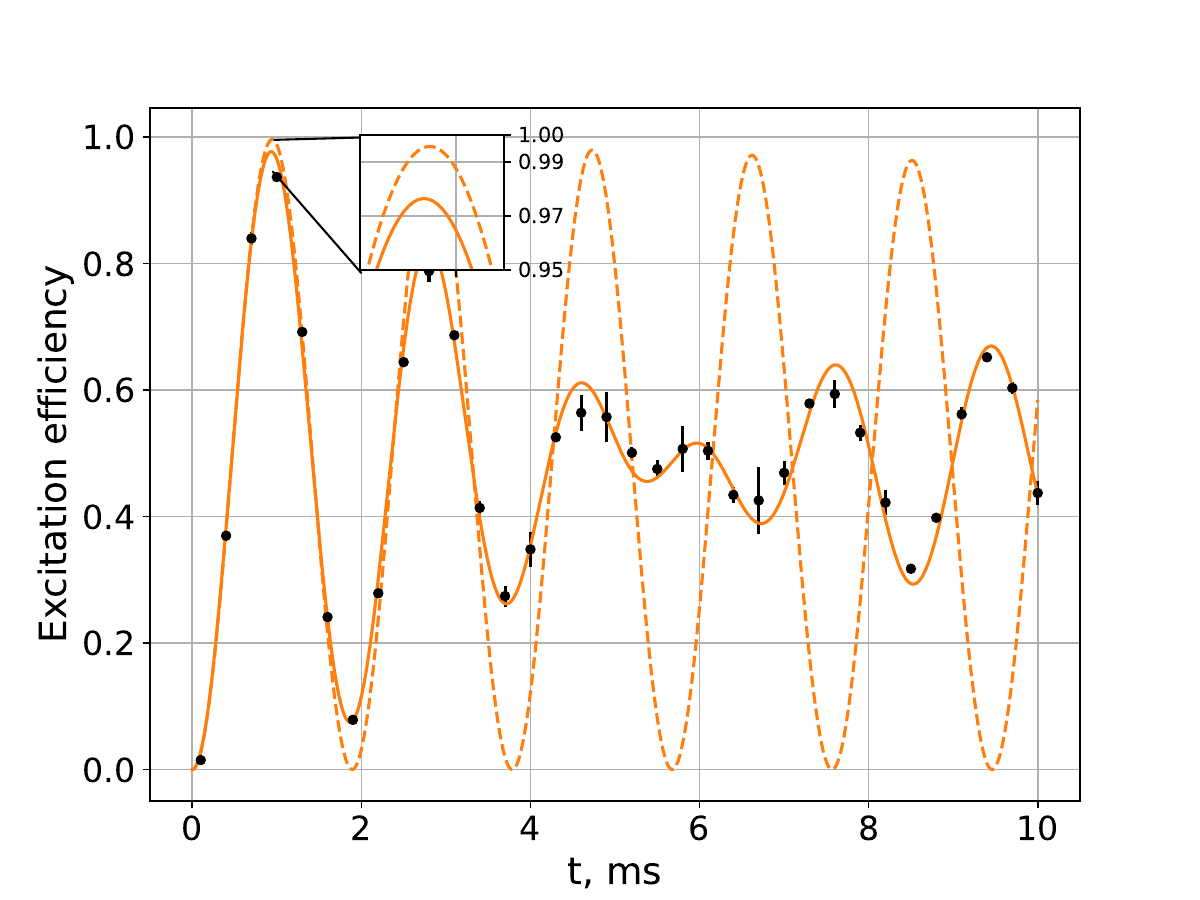}
}\\
\caption{1140\,nm transition excitation efficiency as a function of pulse length.
Error bars correspond to 1\,s.d. statistical uncertainty.
Data is fitted with Eq.\,(\ref{eq:Rabi total}) (solid line) accounting for back reflection from the optical lattice enhancement cavity.
Dashed line illustrates the spontaneous decay limit for the excitation probability without parasitic reflections.
}
\label{fig:Clock Rabi}}
\end{figure*}

To assess the imperfection of excitation caused by the 1140\,nm transition, Rabi oscillations were measured.
For this purpose, the atoms were prepared in the $\ket{g,F=4,m_F=0}$ state, after which, in the readout scheme shown in Fig.~\ref{fig:readout technique}(b), the duration of the first clock pulse was scanned, while the duration of the second pulse, as before, corresponded to a $\pi$-pulse.
The results can be seen in Fig.\,\ref{fig:Clock Rabi}.
The nonmonotonic decay and growth of the oscillation amplitude are attributed to parasitic reflection from the mirror of the enhancing cavity of the optical lattice.
To account for these reflections, a corresponding correction was introduced into the classical Rabi oscillation formula for the population of the excited state $\eta(z,t)$: 
\begin{align}
    \eta(z,t) &= \frac{1-\cos(\Omega(z)\cdot t)}{2}e^{-t/(2\tau_c)},
    \label{eq:Rabi oscillations}
\end{align}
where z is the coordinate along the optical lattice axis and $\tau_c=112$\,ms is the metastable state lifetime.
The Rabi frequency linearly depends on the 1140\,nm radiation field amplitude and for amplitude reflection coefficient $a$ can be expressed as
\begin{equation}
    \Omega(z)=\Omega_0\sqrt{1+a^2+a\,cos(2kz)}.
    \label{eq:Rabi frequency}
\end{equation}
Since the characteristic size of the atomic cloud significantly exceeds the wavelengths of the clock (1140\,nm) and lattice (1063.5\,nm) radiation, averaging over the coordinate can be performed, where for the current model, integration was carried out numerically
\begin{equation}
    \eta(t)=\frac{1}{\lambda}\int_{-\lambda/2}^{\lambda/2}\eta(z,t)\,dz.
    \label{eq:Rabi total}
\end{equation}
Eq.\,(\ref{eq:Rabi total}) was used to fit the obtained data, with the $\Omega_0$ and the reflection coefficient $a$ as free parameters.
In Fig.~\ref{fig:Clock Rabi}, the fit results are shown by a solid line, while the theoretically achievable excitation probability in the absence of parasitic reflections is indicated by a dashed line.

The fit gives an intensity reflection coefficient of about $1.5\,\%$.
We note that this fit does not account for the transverse intensity distribution and radial temperature of atoms, which could be the reason for the imperfect agreement between the experimental data and fitting results. 
Since the wavelength of the clock transition and the magic wavelength for lattice formation are close ($1140$\,nm and $1063.5$\,nm, respectively), designing optics with coatings that provide high reflectivity for one wavelength and low reflectivity for the other is a non-trivial task.
However, this problem does not arise when working with atoms trapped in optical tweezers, as there is no need to use a cavity.
In this case, the efficiency of population transfer will be limited by the lifetime of the metastable level (dashed line in Fig.\,\ref{fig:Clock Rabi}) and the excitation time.
A typical $\pi$-pulse duration for the current system is $1$\,ms, with the possibility of achieving a $50$\,\textmu s $\pi$-pulse without modifying the optical setup.
With tight focusing of the addressing beam (on the order of a few microns) using the same laser system, it is possible to achieve operation speeds of a few microseconds.

Taking all these effects into account allows for the calibration of the readout scheme with an accuracy of approximately $1\%$, which is related to possible fluctuations in the 410\,nm probe pulse power.

\subsection{Coherent preparation of initial states}
\label{Subection:coherent preparation}
\begin{figure*}[ht!]
\center{
\resizebox{0.5\textwidth}{!}{
\includegraphics{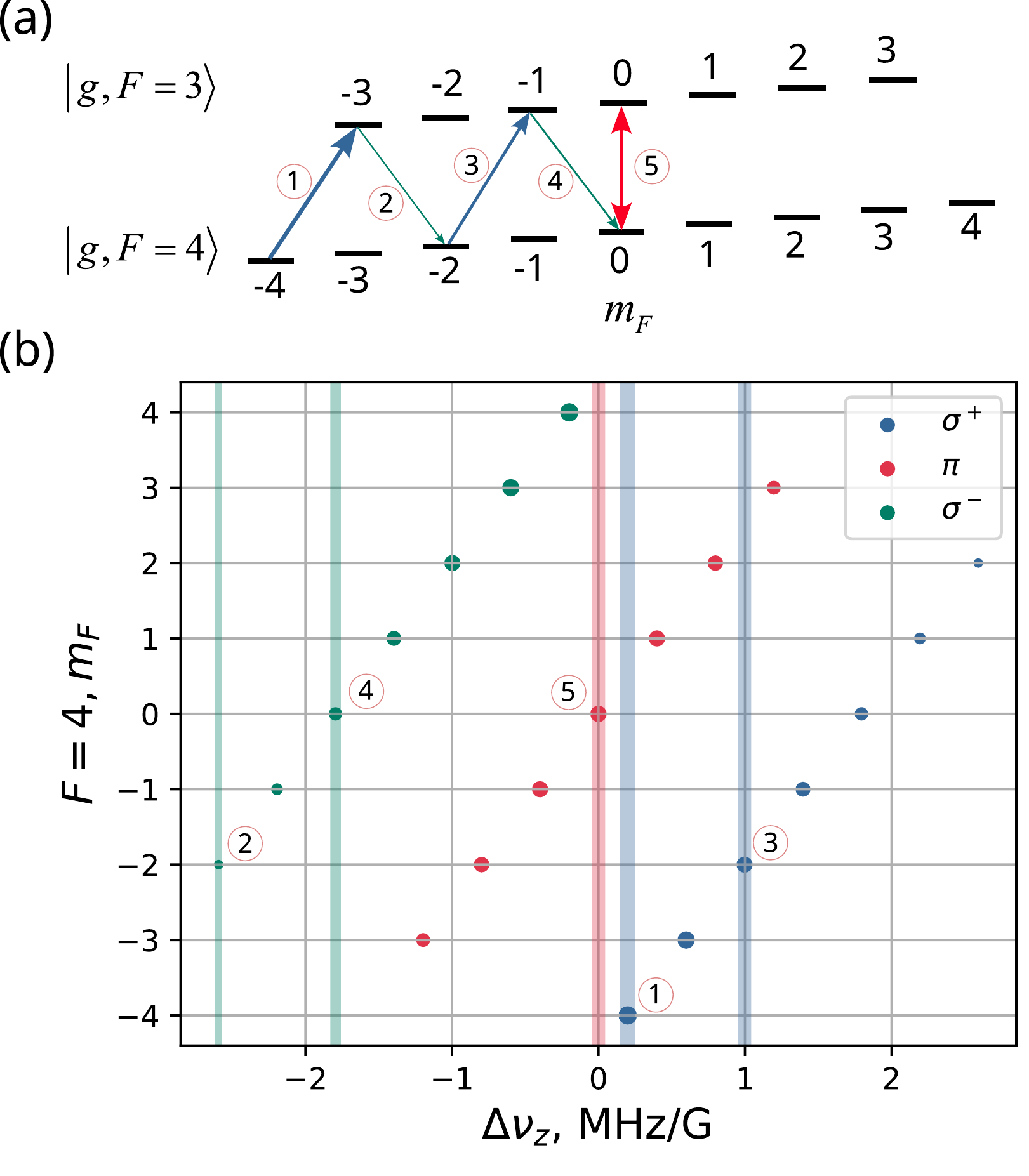}
}\\
\caption{Coherent preparation of the initial state.
Schematic of microwave transitions involved. 
Sigma-polarized transitions 1-4 are used to transfer atoms to the central magnetic sublevel, and transition 5 is used for further cleaning the polarization (see text).
The thickness of the arrows schematically represents the transition strength, corresponding Rabi frequencies relate as $\Omega_1:\Omega_2:\Omega_3:\Omega_4:\Omega_5=1.32:0.25:0.96:0.61:1$.
(b) Linear Zeeman shift coefficient for microwave transitions of different polarizations from different magnetic sublevels of the ground state.
The vertical filled lines mark the transitions that are planned to be used in preparation of the initial states.
The size of the markers and the width of the vertical lines also schematically illustrate the transition strength.
}
\label{fig:coherent repumping}}
\end{figure*}

To improve the efficiency of initial state preparation in further research, coherent population transfer from $\ket{g, F=4, m_F=-4}$ to $\ket{g, F=4, m_F=0}$ state can be utilized.
To achieve this, it is necessary to sequentially excite 4 $\pi$-pulses (or perform rapid adiabatic passage \cite{camparo1984parameters,martin1996coherent}) of microwave radiation corresponding to the transitions indicated in Fig.\,\ref{fig:coherent repumping}.
The minimum frequency splitting between adjacent transitions is on the order of $1.5\,\text{MHz/G}$, allowing all transitions to be excited independently.
Since both strengths and coherence times of these transitions are lower than those of the transition between the central magnetic sublevels considered in this work, implementing this scheme requires a significantly higher microwave radiation intensity and the use of a more complex antenna.

This value can be further improved by increasing the microwave field intensity, stabilizing, and shielding magnetic fields.

When implementing this protocol, non-ideal $\sigma^-$ pulses (pulses (2) and (4) on Fig.\,\ref{fig:coherent repumping}.a) may lead to non-zero population of $\ket{F=3,m_F\ne0}$ sublevels.
To mitigate this, before MW pulse (5) one can implement an additional optical pulse $\ket{g, F=3} \rightarrow \ket{530\,\textrm{nm}, F=4}$ at 530\,nm, detuned by $+614$\,MHz from the second stage cooling transition, to repump atoms from $\ket{g, F=3}$ to $\ket{g, F=4}$.
Then, after transferring atoms from $\ket{g, F=4, m_F=0}$ to $\ket{g, F=3, m_F=0}$, it would be possible to spot ``preparation error'' by detecting the presence of atom in $\ket{g, F=4}$, and if necessary, repeat the initial state preparation.
As an estimate, taking the $98\%$ efficiency of $\ket{F=4,m_F=-4}\rightarrow\ket{F=3,m_F=-3}$ transition demonstrated in \cite{pershin2020microwave}, we get single-try initialization fidelity of $92\%$.

\subsection{Polarization purity of the prepared state}
\label{Subection:polarization_purity}

\begin{figure*}[h!]
\center{
\resizebox{0.5\textwidth}{!}{
\includegraphics{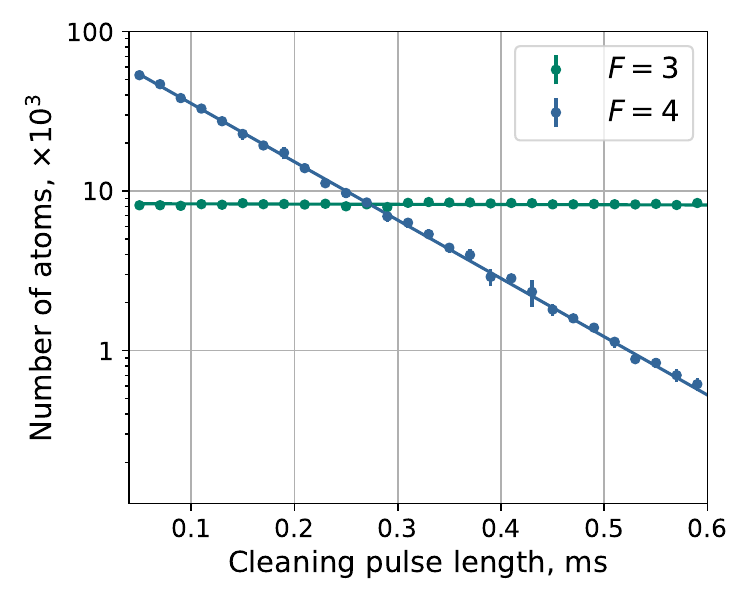}
}\\
\caption{State cleaning.
The dependence of the detected number of atoms in the $\ket{F=3}$ (green points) and $\ket{F=4}$ (blue points) states on the duration of the cleaning pulse at 530\,nm radiation, tuned to resonance with the second-stage cooling transition.
Error bars correspond to 1\,s.d. statistical uncertainty.
The data are fitted with an exponential decay function.
For the $\ket{F=4}$ state, the characteristic time was t = 119(1)\,\textmu s, while no measurable decrease in population was observed for the $\ket{F=3}$ level.
}
\label{fig:cleaning}}
\end{figure*}

During optical pumping, microwave manipulations and cleaning pulse, the following processes may affect the final polarization purity of $\ket{g, F=3, m_F=0}$ state:
\begin{enumerate}
    \item Microwave excitation of $\ket{g, F=3, m_F \ne 0}$ states. 
    For a 2-ms $\pi-$pulse, the Rabi frequency is $\Omega = 2\pi \times 250$\,Hz.
    Since any other microwave transition is detuned by at least $\Delta\nu_\textrm{\tiny{mw}}=60$\,kHz, its maximum excitation probability is $p = \frac{\Omega^2}{\Omega^2 + (2*\pi*\Delta\nu_\textrm{\tiny{mw}})^2} \sim 2\times10^{-5}$.
    \item Some atoms are present in $\ket{g, F=4}$ state after cleaning pulse of resonant 530\,nm radiation. 
    We measure number of atoms remaining in $\ket{g, F=4}$ state as a function of cleaning pulse length, see Fig.\,\ref{fig:cleaning}. 
    The exponential fit gives time constant $\tau_\textrm{\tiny{cl}} = 0.119(1)$\,ms.
    In the experiment, we apply cleaning pulse with duration $t_{c}=3$\,ms, assuring zero atoms population in $\ket{g, F=4}$ state after cleaning pulse.
    \item Depolarization of atoms in $\ket{g, F=3, m_F=0}$ due to scattering of cleaning pulse photons.
    Probability to scatter photon can be estimated as $p \approx \frac{\Gamma_{530} s t_c}{2(1 + s + (4\pi\Delta\nu/\Gamma_{530})^2)} = 3\times 10^{-4}$ for saturation parameter $s=1$, cleaning pulse duration $t_c=3$\,ms, frequency detuning $\Delta\nu=614$\,MHz and $\Gamma_{530}=2\pi\times 350$\,kHz.
\end{enumerate}

Overall, this gives an upper limit on polarization impurity of the initial $\ket{g, F=3, m_F=0}$ state of $p^\textrm{\footnotesize{err}}_\textrm{\footnotesize{tot}} \lesssim 5\times 10^{-4}$.

\subsection{Depolarization}
\label{Subection:Depolarization}

\begin{figure*}[ht!]
\center{
\resizebox{0.9\textwidth}{!}{
\includegraphics{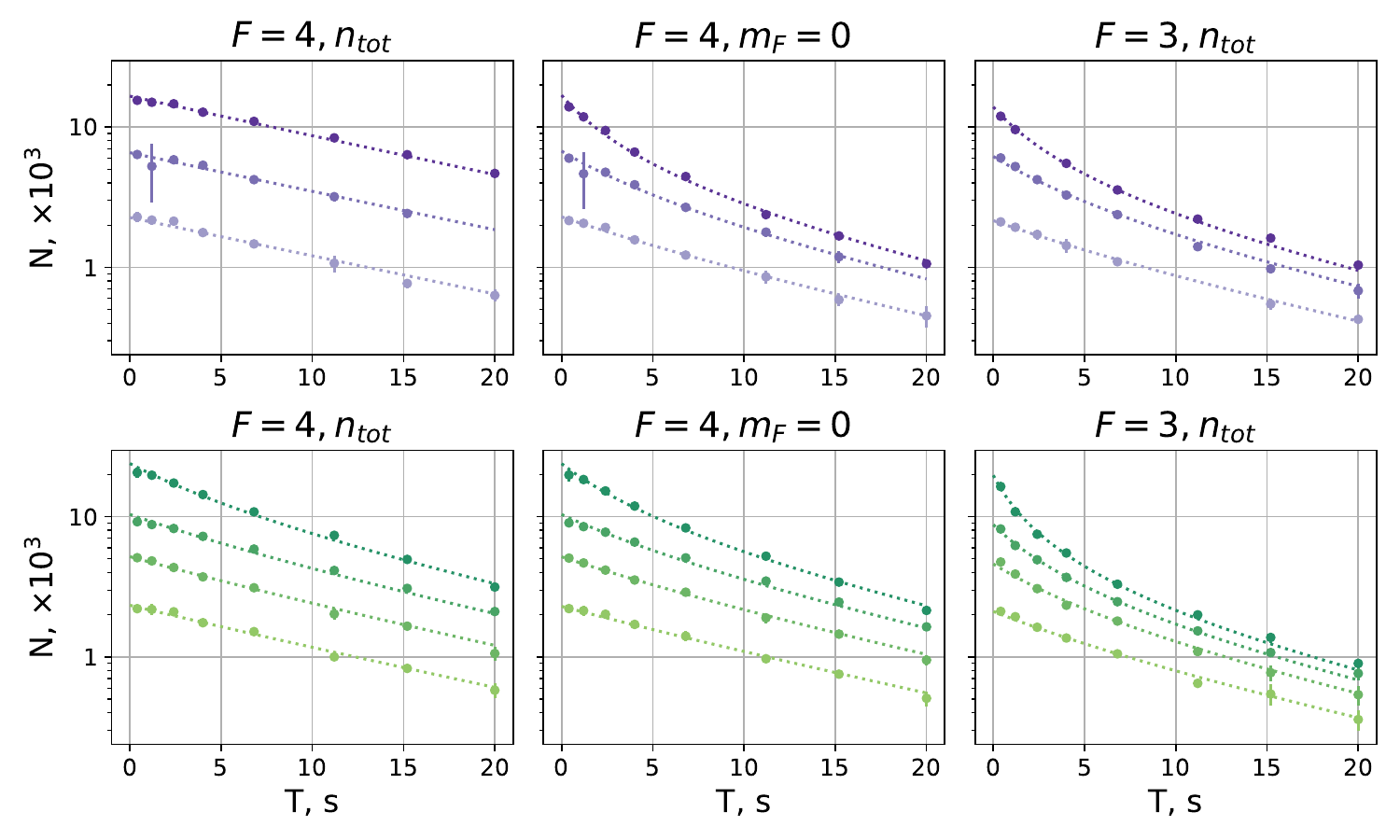}
}\\
\caption{Lifetime and depolarization rate of atoms.
The top row shows the measurement results for a magnetic field of $B=0.1$\,G, and the bottom row shows the results for $B=0.6$\,G.
Error bars correspond to 1\,s.d. statistical uncertainty.
The dependence of the number of atoms on time was fitted with a two-body decay function Eq.~(\ref{eq:two_body_loss}) (dashed line).
At the $\ket{F=3}$ level, the fraction of atoms on the central magnetic sublevel, within the current measurement accuracy, was equal to unity.
}
\label{fig:Depolarization}}
\end{figure*}

The readout scheme described in Sec.~\ref{Section:Readout technique} allows for estimating the depolarization rate of atoms under various experimental parameters.
At time zero, the atoms are prepared in one of the initial states, either $\ket{F=4,m_F=0}$ or $\ket{F=3,m_F=0}$.
After a waiting period $T$, the readout procedure is performed.
Fig.\,\ref{fig:Depolarization} presents the measurement results: different shades of purple and green correspond to different initial numbers of atoms in the experiments with $B=0.1$\,G and $B=0.6$\,G, respectively.
The number of atoms was fitted with the two-body decay function Eq.\,(\ref{eq:two_body_loss}) (dashed lines).
For the $\ket{F=4}$ level, we present the measurement results of the total number of atoms and the number of atoms in the central magnetic sublevel.
For the $\ket{F=3}$ state, only the graph of the total number of atoms as a function of time is provided, as no depolarization was observed, and within the current measurement accuracy, the population fraction of the central magnetic sublevel $\ket{F=3,m_F=0}$ remained constant at unity throughout the entire experiment.

For the total atom number in the $\ket{F=4}$ state $n_{tot}=n_{m_F=0}+n_{m_F\ne0}$ for both magnetic fields, the parameter $\tau=16(2)$\,s closely matches the lifetime measurement results for the lowest energy state $\ket{F=4,m_F=-4}$, described in the main text, which validates our choice of a fixed $\tau=16.4$\,s value for fitting these data.
Similarly to the lowest energy level lifetime analysis, at higher magnetic fields, the two-body interaction effects become more pronounced: $\beta(F=4,n_{tot},B=0.1\,G)=3.2(3.4)\times10^{-11}\,\textrm{cm}^3/\textrm{s}$ and $\beta(F=4,n_{tot},B=0.6\,G)=6.2(2.1)\times10^{-10}\,\textrm{cm}^3/\textrm{s}$, which can also be attributed to the proximity to a Feshbach resonance \cite{khlebnikov2019random}.

Comparing to the $\ket{F=4,m_F=-4}$, state, lifetime analysis of the central magnetic sublevels reveals a significantly stronger influence of two-body interactions.
The corresponding coefficients of the order of $3\times10^{-9}\,\textrm{cm}^3/\textrm{s}$ were already presented in Table.\,\ref{tab:lifetime}.
We observe an asymmetry between the F=4 and F=3 states:
\begin{itemize}
    \item Atoms undergoing loss from the $\ket{F=3, m_F=0}$ level leave the optical lattice 
    \item Atoms in the $\ket{F=4, m_F=0}$ state undergo depolarization through redistribution across other magnetic sublevels without leaving the trap
\end{itemize}
We attribute this behavior to dipole-dipole interactions (DDI): interatomic interactions can induce spin flips, while in the ground state the released energy remains insufficient for atoms to escape the optical lattice potential (in all experiments, approximately $U=k_B \times5\,\mu \textrm{K}=100\,E_{\textrm{rec}}$).
However, for the $F=3$ level, additional decay channels into the ground state become possible with potential energy release a few orders higher than optical lattice depth.
The readout scheme developed in this work enables selective population measurements of central magnetic sublevels in all conducted experiments, thereby accounting for depolarization effects.
This method can be applied to any study or protocol where the metastable-state population at the readout moment can be considered negligible.
However, we expect that for individually trapped atoms in optical tweezers, both DDI and depolarization will be suppressed to a level that does not compromise operational fidelity.

\subsection{Ramsey fringes analysis}
\label{Subection:Ramsey fringes analysis}

\begin{figure*}[ht!]
\center{
\resizebox{\textwidth}{!}{
\includegraphics{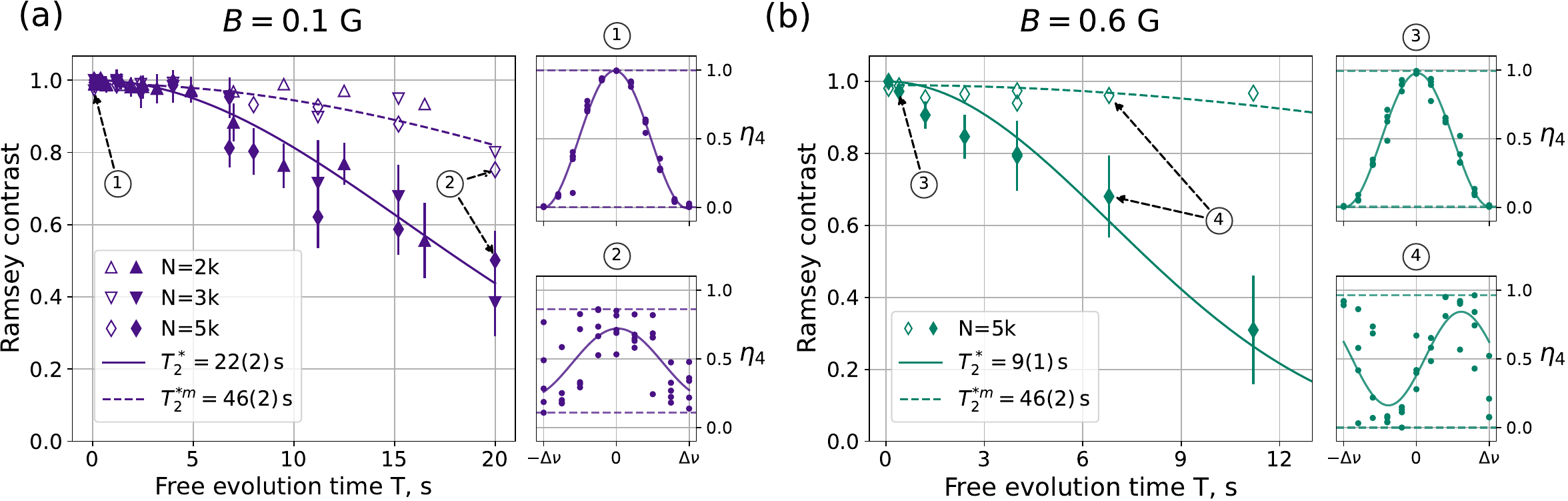}
}\\
\caption{
Ramsey fringes analysis.
The subplots (numbered 1-4) on the graphs illustrate the Ramsey fringes measurements for selected free evolution times T, as annotated.
On these subplots $\Delta\nu=1/T$, dashed horizontal lines show maximum and minimum probabilities detected and solid lines show the fit with Eq.\,(\ref{eq:Ramsey fringes}).
Solid markers with error bars on the main graphs represent the contrast and 1\,s.d. obtained from such fits, while empty markers correspond to the estimate based on the difference between the maximum and minimum measured probabilities.
Both datasets were fitted with a Gaussian decay function Eq.\,(\ref{eq: gaussian decay}). 
For the filled markers, the characteristic decay times were $T_2^*$ of $22(2)$\,s and $9(1)$\,s for $B=0.1$\,G and $B=0.6$\,G respectively (solid lines).
For the empty markers at $B=0.1$\,G, the estimated decay time was $T_2^{*m}=46(2)$\,s, which also provides reasonably good description of the behavior at $B=0.6$\,G (dashed lines).
}

\label{fig:Ramsey_fringes}}
\end{figure*}

Here, we present a more detailed analysis of Ramsey spectroscopy results.  
In Fig.\,\ref{fig:Ramsey_fringes}, the insets show Ramsey spectroscopy results for different free evolution times, as labeled with numbers 1-4.
The experimental data in each case were fitted with Eq.\,(\ref{eq:Ramsey fringes}) (solid curves), and solid markers in the main Fig.\,\ref{fig:Ramsey_fringes}(a,b) illustrate Ramsey contrast achieved from this fit.
As one can see from the insets with short free evolution times (1 and 3) the shot-to-shot fluctuation of measured $\ket{g,F=4, m_F=0}$ fraction of atoms $\eta_4$ is small, and individual data points are well aligned with the fit.
At longer times (insets 2 and 4), scattering of individual $\eta_4$ points measured at the same frequency detuning becomes larger, reaching $\sim50\%$ for the chosen $T$. 
This behavior can be interpreted as the transition frequency drifts between measurements, which can be associated with slow (1\,s or slower) magnetic field fluctuations.
The observed scattering of consecutive measurements for both $B=0.1$\,G and $B=0.6$\,G can be explained by magnetic field fluctuations at a level below $150\,$\textmu G, which matches our previous estimations \cite{mishin2022effect}.

In the presence of such fluctuations, the decoherence time can be estimated not from fitting-derived data (which are affected by noise from temporal field fluctuations, yielding a lower-bound estimate), but rather from the difference between maximum and minimum detected probabilities \cite{kleine2011extended}.
Here we assume that temporal magnetic field fluctuations cannot compensate for the atomic cloud decoherence process.
Dashed horizontal lines in the insets illustrate minimum and maximum values measured in the corresponding experiment.
Empty markers in Fig.\,\ref{fig:Ramsey_fringes} depict the contrast estimated from the difference between maximum and minimum probabilities measured during the corresponding free evolution time.

All measured Ramsey contrast dependencies are fitted with a Gaussian decay function Eq.\,\ref{eq: gaussian decay}.
For data extracted from Ramsey oscillation fits, the corresponding coherence times $T_2^*$ are $22(2)$\,s and $9(1)$\,s for $B=0.1$\,G and $B=0.6$\,G respectively, depicted with solid lines in Fig.\,\ref{fig:Ramsey_fringes})(a,b).
For the peak-to-peak probability estimates at $B=0.6$\,G, no significant decay is observed due to the shorter experimental timescale; nevertheless, these data remain well-described by the $T_2^{*m}=46(2)$\,s derived from fitting analogous $B=0.1$\,G results, depicted with dashed lines in Fig.\,\ref{fig:Ramsey_fringes})(a,b)

The peak-to-peak data analysis yields an upper bound of $\sim50\,$\textmu G, for the magnetic field inhomogeneity across the atomic cloud, with a coherence decay time constant consistent with that obtained from dynamical decoupling measurements.
This agreement further supports magnetic field fluctuations as the dominant decoherence mechanism in the present experimental configuration.

\subsection{Dynamical decoupling measurements}
\label{Subection:Dynamical decoupling measurements}

In the current experimental setup, it was not possible to directly adjust the phase of microwave pulses relative to one another, so we used the following reasoning to estimate the contrast in the dynamical decoupling experiments:
\begin{enumerate}
    \item Since the developed readout scheme allows us to separately address atoms in the central magnetic sublevels, and the decay from $F=3$ to $F=4$ level leads to loss of an atom, we expect the oscillations (if scanning the phase of the last microwave pulse) to decay symmetrically over time converging to flat line at $1/2$.
    As a result, the contrast of the oscillations will tend to zero, while the probability of finding the system in either the $F=3$ or $F=4$ state will approach $1/2$.
    This assumption is supported by the results of Ramsey spectroscopy conducted earlier: for all recorded oscillation spectra, the offset $A$ in Eq.~(\ref{eq:Ramsey fringes}) was equal to $0.5$ within the fitting error margin.
    In the experiment, we aim to measure the maximum probability $\eta_{max}$ of finding the system in the target state  (which depends on the initial state and the number of intermediate $\pi$-pulses), i.e., when the relative phase of the final pulse is zero.
    The $\eta_{max}$ dependence on time in this approximation can be described by
    \begin{equation}
    \eta_{\textrm{max}}(t)=\frac{\eta_{\textrm{max}}(0)}{2}e^{-\left(t/T\right)^2} +\frac{1}{2}
    \label{eq: gaussian decay offset}
    \end{equation}
    
    \item The obtained values will represent a lower bound estimate, since magnetic field fluctuations and a number of other effects can disrupt the phase, causing the system to deviate from the excitation peak, while pure decoherence leads to a uniform decay of oscillations and should not vary from measurement to measurement. 
\end{enumerate}
In these approximations, the contrast can be estimated using the obtained probability $\eta_{max}$ of finding the system in the target state via the formula
\begin{equation}
    C=1-2(1-\eta_{\textrm{max}})=2\eta_{\textrm{max}}-1
    \label{eq: decoupling contrast}
\end{equation}

\begin{figure*}[ht!]
\center{
\resizebox{1\textwidth}{!}{
\includegraphics{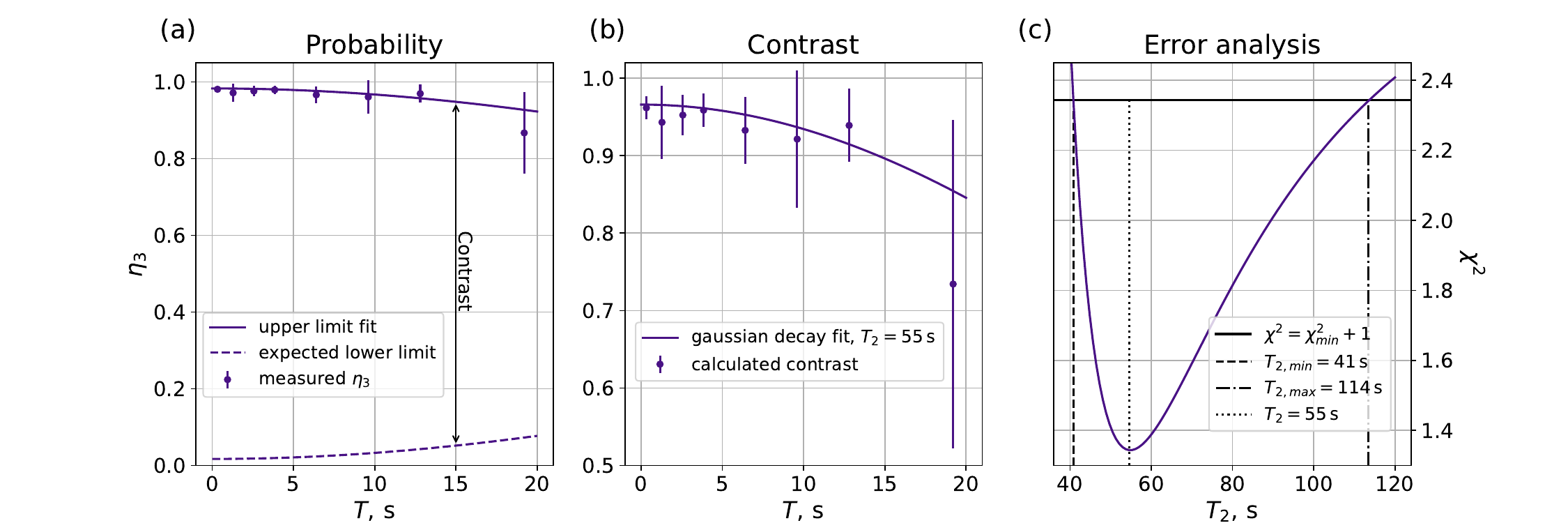}
}\\
\caption{
An example of processing the results of dynamic decoupling for $B=0.1$\,G and $n=8$ intermediate pulses.
(a) The experimental points with error bars correspond to the detected probability and 1\,s.d. statistical uncertainty of finding the system in the $\ket{F=3}$ state.
The solid line represents the fit of these results with Eq.\,(\ref{eq: gaussian decay offset}), and the dashed line corresponds to the symmetric decay of the lower probability limit and is provided to illustrate the calculation of the contrast from the obtained values.
(b) Inferred time dependence of contrast is fitted with gaussian decay function Eq.\,(\ref{eq: gaussian decay}), similar to Fig.\,\ref{fig:Decoupling}.
We used the $\chi^2$ parameter to estimate the uncertainty in determining $T_2$, as shown in the figure (c). 
}
\label{fig:Decoupling processing}}
\end{figure*}

An example, the  processing of the experimental data for $B=0.1$\,G and $n=8$ is shown in Fig.~\ref{fig:Decoupling processing}(a).
For these conditions, the target state is $\ket{F=3}$, and the experimental dots illustrate the probability of finding the system in this state. 
The solid line represents the fit with Eq.~(\ref{eq: gaussian decay offset}), and the dashed line corresponds to the symmetric decay of the minimum probability and is provided to illustrate the calculation of the contrast from the obtained $\eta_3$ values.
The inferred contrast is shown on Fig.~\ref{fig:Decoupling processing}(b). 
To evaluate the error of the obtained coherence time, the $\chi^2$ parameter was used, as shown in Fig.~\ref{fig:Decoupling processing}(c).

\section{Acknowledgments}

The authors acknowledge the support of RSF grant no. 24-72-10102. 

\bibliography{biblio}

\end{document}